%% file: main.tex
\documentclass[10pt]{iopart}
\usepackage{iopams}
\usepackage{multicol}
\usepackage{multirow}
\usepackage{siunitx}
\input{style/flowchart_settings}
\input{style/matlab_code_settings}

\renewcommand{\theenumi}{\arabic{enumi}.}

\renewcommand{\theenumii}{\theenumi\arabic{enumii}}

\begin{document}
\ioptwocol[\title{Feature Characterization for Profile Surface Texture}
\author{A Müller $^1$, M Eifler$^1$, A Jawaid$^1$, J Seewig$^1$}
\address{$^1$ RPTU Kaiserslautern-Landau, Germany}
\ead{alexander.mueller@mv.rptu.de}
\begin{abstract}
Conventional field parameters for surface measurement use all data points, while feature characterization focuses on subsets extracted by watershed segmentation. This approach enables the extraction of specific features that are potentially responsible for the function of the surface or are a direct reflection of the manufacturing process, allowing for a more accurate assessment of both aspects. Feature characterization with the underlying watershed segmentation for areal surface topographies has been standardized for over a decade and is well established in industry and research. In contrast, feature characterization for surface profiles has been standardized recently, and the corresponding standard for watershed segmentation is planned to be published in the near future. Since the standards do not provide guidelines for implementation, this paper presents an unambiguous algorithm of the watershed segmentation and the feature characterization for surface profiles. This framework provides the basis for future work, mainly investigating the relationship between feature parameters based on feature characterization and the function of the surface or manufacturing process. For this purpose, recommendations for the configuration and extensions of the toolbox can also be developed, which could find their way into the ISO standards.
\end{abstract}
\noindent{\it Keywords\/}: profile surface texture, feature characterization, watershed segmentation]

\section{Introduction} \label{introduction}
Field parameters condense all data points of a surface measurement into a single parameter or set of parameters using mathematical rules. However, certain surface properties are significantly determined by distinct features of the surface, such as the hills respectively peaks of a surface in tribological contact. The approach of feature characterization is therefore to extract the relevant parts by watershed segmentation of the surface and to derive so-called feature parameters from them. For industrial surfaces, feature characterization for areal surface measurement was first investigated in 1993 \cite{Zahouani.1993} and has been standardized in ISO 25178-2 \cite{25178-2} since 2012. The full history can be found at \cite{Blateyron2013}. The associated watershed segmentation inspired by geodesy was moved to ISO 16610-85 as a filter one year later. Since the method has been standardized, multiple applications have emerged, which can be divided into two groups. The first group are parameters related to the function of the surface, and are therefore used to evaluate the part itself in its given application. For example, it has been shown in various publications that the peaks of a surface are decisive for wear \cite{Zabala_2020,TIAN_2011,TIAN_2012,Dimkovski_2018} or influence the shear strength of adhesive bonds \cite{ZIELECKI_2013}. In \cite{Stöckel_2023}, feature characterization is used as a basis to assess the spatial distribution of features. The second group are parameters that are related to the manufacturing process and can be used for process monitoring or fault diagnosis. Particularly in additive manufacturing, feature parameters are used to assess the surface topography \cite{Senin_2018,MACULOTTI2022,LOPEZ_2023,LOPEZ2023_2}. But also other manufacturing processes such as injection molded car interior parts \cite{Reddy_2023}, laser-assisted machining \cite{PU_2020}, coatings for photovoltaic substrates \cite{Blunt_2014} or in the manufacture of microelectromechanical systems/microoptoelectromechanical systems (MEMS/MOEMS) \cite{Osten.2020} are evaluated with feature characterization. It is also used to evaluate the manufacturing tool itself \cite{ISMAIL2011,YE2016}.

Nevertheless, there are suggestions for improving the areal feature characterization according to ISO 25178-2 \cite{25178-2}. In \cite{Wolf.2020,Wolf.2021}, Wolf examines the connections to Morse theory and highlights the advantages of transferring feature characterization and suggests adapting the terms from ISO 25178-2 \cite{25178-2} accordingly. However, it is not yet foreseeable whether and when corresponding changes will be made to ISO 25178-2 \cite{25178-2} and 21920-2 \cite{21920-2}, so only the existing standards are used here.

\begin{figure*}
    \centering
    \includegraphics[width=\textwidth]{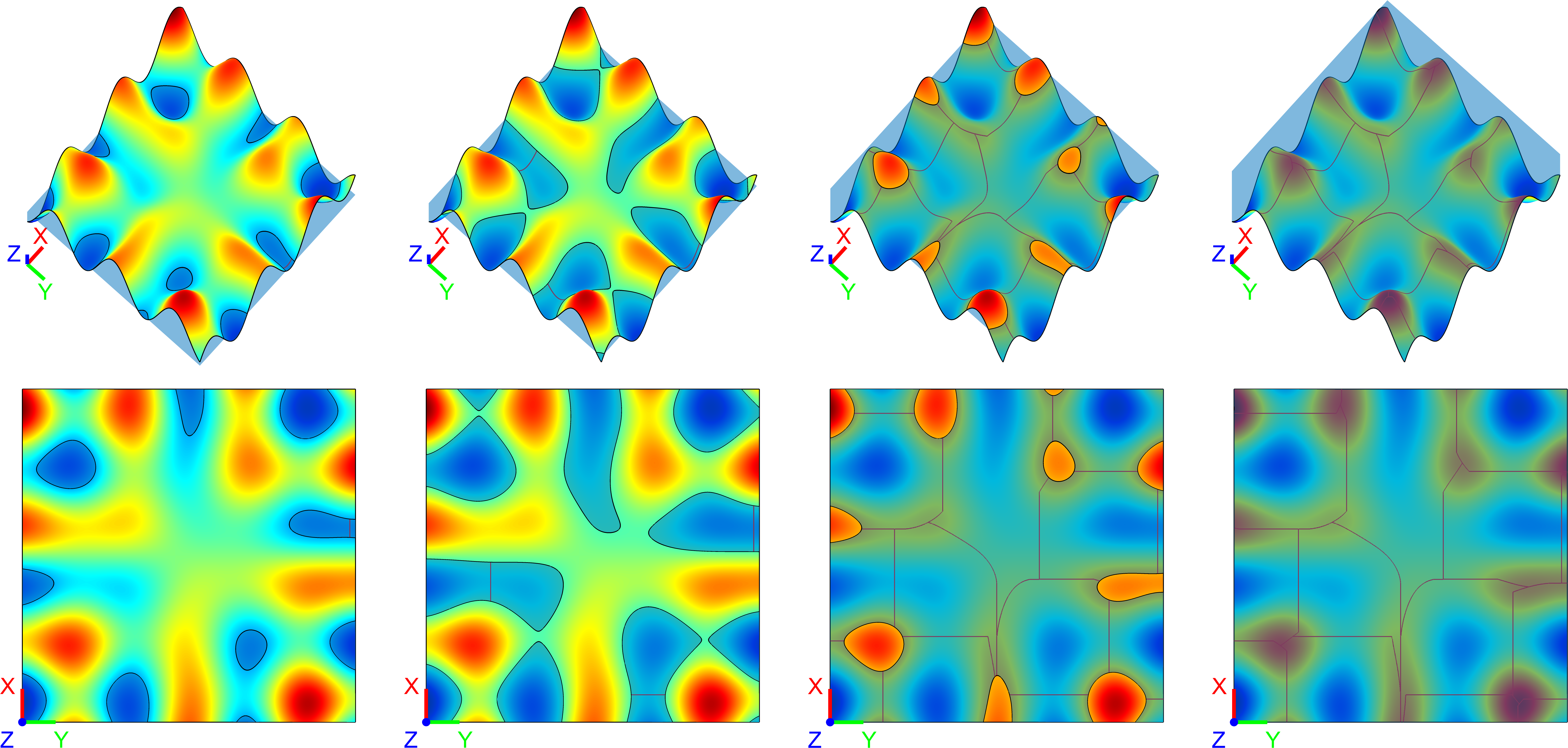}
    \caption{Illustration of the principle of watershed segmentation on an areal topography with rising water level from left to right. The red lines represent the watersheds.}
    \label{visu_3D}
\end{figure*}
For profile surface measurement, feature characterization has been transferred to the ISO 21920-2 standard \cite{21920-2} in 2021. A distinction is made here between feature parameters based on peak heights and pit depths, feature parameters based on profile elements and feature parameters based on feature characterization. The latter are equivalent to areal feature parameters, which are also based on watershed segmentation (see section \ref{watershed_sec}). An alternative approach to profile segmentation is the motif method, which is standardized in ISO 12085 \cite{12085}. The features, also called motifs, are determined here from the sequences of peaks and pits, from which the so-called R\&W parameters are derived. These parameters are mainly used in the French automotive industry. Blateyron~\cite{Blateyron_2021} made a comprehensive comparison between the segmentation methods, which shows that watershed segmentation is more stable and robust. In addition, he was also able to show under which configuration the corresponding parameters correlate strongly.

In summary, areal feature characterization is a well-established method with the capacity to assess the function and the manufacturing of surfaces. This approach can now be extended to profiles, a process that has already been initiated by ISO 21920-2 \cite{21920-2} and ISO 16610-45 \cite{16610-45}. The purpose of this paper is to present a clear and detailed implementation guideline for the feature characterization for profiles, which is currently missing.
Some corrections and extensions are necessary. One of the primary reasons for this is that the standard refers to continuous profiles, whereas the practical measurement data that is generally presented as a digital signal, which requires a discrete implementation. 
%

The algorithm is divided into two main functions. First, the features are determined using watershed segmentation (see section 2). Second, feature parameters are derived from the features obtained (see section 3). On the one hand, this allows the watershed segmentation to be carried out separately and the settings to be adjusted if necessary. On the other hand, different parameters can be derived from the obtained features without having to redetermine the features for each parameter. With the main function feature characterization (see section \ref{FC_sec}), both watershed segmentation and feature parameters can be executed in one step. In order to maintain readability and comprehensibility, only the main steps are described, which can also serve to highlight the characteristics of the method. Minor help functions, some of which are required in several places, have been moved to the appendix.
Finally, the plausibility of the algorithm and possible applications are shown using example profiles (see section \ref{example_profiles}). 
\section{Watershed segmentation on profiles} \label{watershed_sec}
Watershed segmentation has its origins in geodesy and was first described by Maxwell more than 150 years ago \cite{Maxwell.1870}. He was able to prove that a landscape can be characterized by point, line, and area elements. The principle can also be applied to areal or even 3D datasets in other fields of applications. In particular, it is used in digital image processing as a segmentation method and is considered a basic element of object recognition with a wide range of applications such as medicine, materials science or quality control.
\begin{figure}
    \centering
    \includegraphics[scale=1]{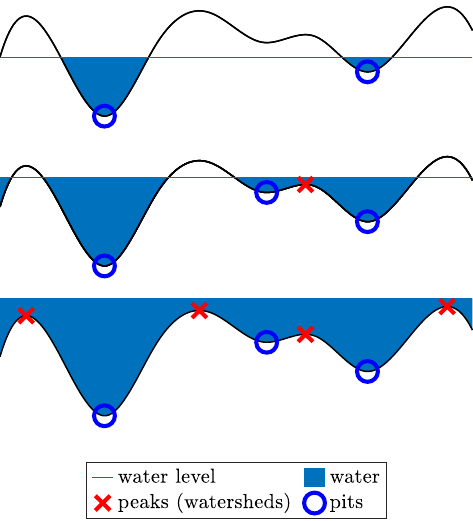}
    \caption{Illustration of the principle of watershed segmentation on a profile with rising water level from top to bottom.}
    \label{visu_2D}
\end{figure}
In the context of surface texture measurement, watershed segmentation was already mentioned by Blunt in 2003 \cite{Blunt.2003} and was included for the first time in 2012 in ISO 25178-2 \cite{25178-2} and moved to ISO 16610-85 \cite{16610-85} as a filter one year later. In the standard, watershed segmentation is applied to areal topographies to divide the surface into segments and derive corresponding parameters. The application of watershed segmentation to profiles is already outlined in ISO 21920-2 \cite{21920-2} with corresponding definitions and will be described in detail in the filter standard ISO 16610-45 \cite{16610-45}, which is currently in the stage of final draft international standard (FDIS).

Watersheds occur in nature. The most famous example is the watershed in North America that separates the Atlantic and the Pacific Ocean, and in this context describes the separation between two neighboring drainage systems. A raindrop falling on the side facing the Atlantic will theoretically, sooner or later, flow into the Atlantic. The same applies to the side facing the Pacific. The boundary line between the two regions is called the watershed. A simple method for determining watersheds is shown in Figure \ref{visu_3D}. It goes back to Vincent and Soille \cite{Vincent.1991}, who published an efficient algorithm for digital image processing, based on which most of the today's algorithms were developed. The idea behind this method is to immerse a water-permeable surface in water. When the surface is submerged, collection basins are formed in the valley bottoms that grow larger as the water level rises. Along the way, water from different basins meets. These meeting points define the watersheds. This allows a complete subdivision (segmentation) of the surface into valleys. The surface can also be segmented into hills. The equivalent idea would be to fill the profile with rising steam from below. The collection points of the steam correspond to the hills. The contact points of the different collection points are again the watersheds. However, it is equivalent to mirroring the topography on the xy-plane, so it can be segmented by dales, as described above.

This method can also be applied to a profile as shown in figure \ref{visu_2D}. In this case, the watersheds are point-like and reflect the peaks (local maxima) of the profile. The watersheds of a profile do not necessarily correspond to the watersheds of the areal topography. For example, a drop hitting a hill could flow into a dale orthogonal to the cross-section, which was not captured in the section. When evaluating profiles, however, it is recommended that profiles perpendicular to the preferred direction of anisotropic or at least semi-anisotropic surfaces (e.g., turned or ground surfaces) be used, as these profiles are representative of the entire surface. For such profiles, the watersheds of the profile most likely correspond to those of the areal topography.

\subsection{Terms and definitions} \label{definitions}

\begin{figure*}
    \centering
    \includegraphics[scale = 1]{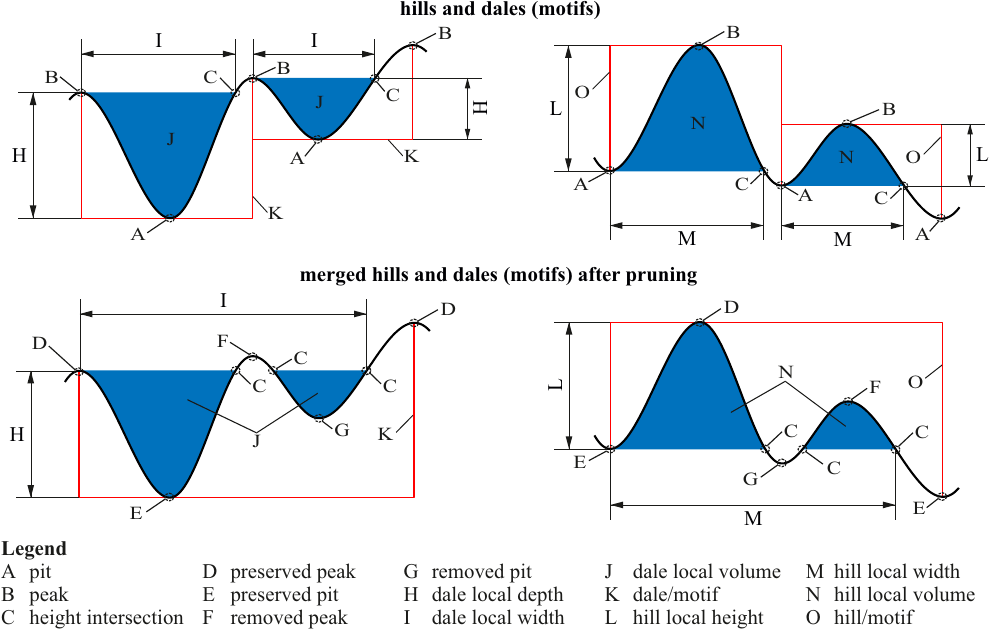}
    \caption{Convention for naming features and their attributes. A special case is presented at the bottom to illustrate how to deal with a removed peak (pit) that lies above the lower peak (below the higher pit) of the motif.}
    \label{conv}
\end{figure*}

In the following, the definitions of the individual terms from ISO 21920-2 \cite{21920-2} are used, which are the initial point for the presented implementation. Some definitions are extended because the standard assumes a continuous representation of the profile, which leads to gaps in the handling of the actual discrete data sets. Figure \ref{conv} shows a dale (left) and a hill (right), and some attributes can be derived from them. Later on, and also in the presented algorithm itself, when hill features are searched, the profile is mirrored on the x-axis and segmented according to the dales. Consequently, the following descriptions are pertaining exclusively to the dales.
\begin{table*}
\centering
\caption{In- and output variables for the watershed segmentation} \label{IO_variables}
\begin{tabular}{| p{0.48cm} p{7.22cm}  p{0.62cm} p{7.08cm}|}
\hline
\multicolumn{2}{|l}{\textbf{Input variables}}& \multicolumn{2}{l|}{\textbf{Output variables}}\\
\hline
$\bi{z}$ &  \multirow[t]{2}{7.2cm}{vector of the ordinate values of the profile, where $z_k$ represents the individual values for $k = 1, 2, \ldots, n$ and $n$ is the number of profile values.} &
${M}_k$       & motif with five members with ${k=\{1,2,..., n_\mathrm{M}\}}$ where $n_\mathrm{M}$ is the total number of motifs\\
& & ${M}.i_\mathrm{v}$     & interpolated index of the pit of dale (peak of hill)\\
$dx$        & sampling interval &  ${M}.i_{\mathrm{lp}}$  & \multirow[t]{2}{7.2cm}{interpolated index of the lower peak of dale (higher pit of hill)}\\
$FT$        & feature type $\in$ \{$"\mathrm{D}"$, $"\mathrm{V}"$, $"\mathrm{H}"$, $"\mathrm{P}"$\} & & \\
$PT$        & pruning type $\in$ \{$"\mathrm{None}"$, $"\mathrm{Wolfprune}"$, $"\mathrm{Width}"$, $"\mathrm{VolS}"$, $"\mathrm{DevLength}"$\} & ${M}.i_{\mathrm{hp}}$  & interpolated index of the higher peak of dale (lower pit of hill)\\
$TH$        & threshold-value for pruning $\in \mathbb{R}^+_0$ & ${M}.sig$     & indicator for significant motifs (1: significant, 0: not significant)\\
\hline
\end{tabular}
\end{table*}
A peak (see key B in figure \ref{conv}) is defined as the highest point within an environment. The lowest point within an environment is called a pit (see key A in figure \ref{conv}). There is a theoretical possibility that a pit or peak may occur as a plateau. In this case, the center of the plateau is assumed to be the pit or peak. Since we are assuming a discrete data set, it is possible for the center of a plateau to lie between two individual points. Therefore, linear interpolation is used to determine the center position. The generic term for peak and pit is point feature, which can be removed by so-called pruning (see below) to merge segments that are too small with their neighbored segments, which is relevant for the next definitions. A dale (see key K in figure \ref{conv}) is the region around a pit where all maximum downward paths end in the pit. The removed peaks and pits are ignored within the dale (see figure \ref{conv}). A dale is surrounded by two adjacent preserved peaks. A dale or a hill can both be called line features or motifs.

One challenge in watershed segmentation is usually oversegmentation, i.e., the identification of numerous small segments that are actually part of larger structural units. Two major factors contribute to this phenomenon. First, measurement errors and noise are usually present when the surface data is acquired. Second, defects in the surface caused by wear or during the manufacturing process can also contribute to oversegmentation. This is where the above-mentioned pruning comes into play, where too small motifs are merged with their neighbored motif. The smallest element is determined according to various size criteria (see attributes below). If its attribute value is lower than a selected threshold, the element is merged with one of its neighboring motifs. The dale into which the water would overflow first if the dale to be merged were to flood further is used for merging. The pit and the lower peak of the too small dale are then removed (see figure \ref{conv}). If both peaks of the dale have the same height, the left neighbor is selected. Even though this case will rarely occur in practice, it should be noted that a mirrored profile would give different results. Several attributes can be derived from the obtained motifs, as shown in figure \ref{conv}. 

The dale local depth (see key H in figure \ref{conv}) describes the height difference between the preserved pit and the preserved lower peak of a dale. 

For the other two attributes shown in the figure \ref{conv}, the height intersections (see key C in figure \ref{conv}) are relevant, which is not explicitly defined in ISO 21920-2 \cite{21920-2} but is the prerequisite for determining several attributes. It describes the intersection with the dale with the height of the lower peak. Figure \ref{conv} below shows a special case where the removed peak of the pruned dale lies above the lower peak. This means that there can be several height intersections per motif. This case is not explicitly addressed in ISO 21920-2 \cite{21920-2}, and in the following we suggest how to deal with such cases for the relevant attributes. 

The dale local width (see key I in figure \ref{conv}) is defined in ISO 21920-2 \cite{21920-2} as “length of the line intersecting a dale at height associated to the lowest peak connected to that peak”. In the case shown, the interruption caused by the removed peak of the dale would therefore be subtracted. However, since the removed peak is considered an error or artifact due to oversegmentation, it should not be considered in the width of the motif. The definition of dale local width used in this algorithm is therefore: “horizontal distance between the preserved lower peak of a dale and the first intersection with the height associated to that lower peak outgoing from the preserved higher peak of the dale”.

The dale local volume (see key J in figure \ref{conv}) describes the “ratio of the dale area below the lowest peak connected to that dale to evaluation length” \cite{21920-2}. This area is marked as blue in figure \ref{conv}. It is expressed in milliliters per square meter ($\mathrm{ml/m^2}$). As in the case of figure \ref{conv} below, the area can be divided into several sub-areas. All height intersections are required to determine the area. 

In addition to these illustrated attributes, there are dale local developed length, peak height, curvature, and count.
The dale local developed length is listed as an attribute in the annex of ISO 21920-2 \cite{21920-2}, but an explicit definition is missing. In this publication, it is defined as the path length between the preserved lower peak of the dale to the first height intersection outgoing from the preserved higher peak. 

The pit depth is the depth difference between the preserved pit of a dale to the reference line.

The local curvature describes the curvature of the preserved pit of a dale.

With $"\mathrm{Count}"$ the attribute value of the line/point element takes a value of one which allows counting the line or point elements, for example, to determine the density of peaks.
\begin{figure}
    \centering
    \includegraphics[scale=1]{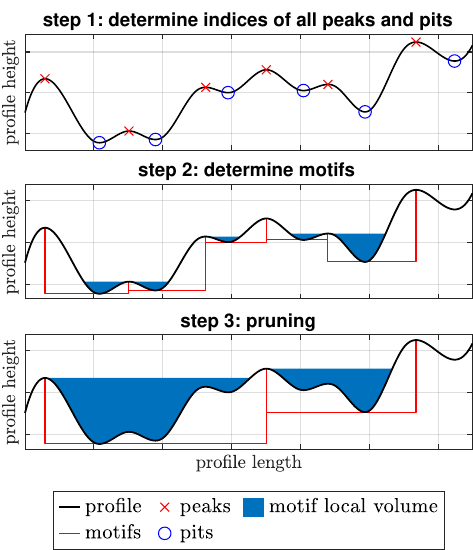}
    \caption{Three steps of watershed segmentation of 2D roughness profiles.}
    \label{steps}
\end{figure}

Accordingly, to describe a dale, the position of the preserved pit, the position of the preserved lower peak and the position of the preserved higher peak are required. For certain attributes, and not least for visualization purposes, the positions of the height intersections are also stored. As mentioned above, a peak or pit in case of a plateau can lie between two discrete points and the positions are therefore interpolated linearly. The same applies to the height intersections. Since the algorithm requires an equidistant sampled roughness profile, the positions can be interpolated using the indices of the height points. In this way, all positions of the points mentioned can be saved as (interpolated) indices with which both $z$ and $x$ coordinates can be retrieved as required. Furthermore, a motif may be designated as non-significant at a later point. To accommodate this possibility, another variable, $sig$, is introduced. This variable takes on the values $1$ and $0$, respectively, to indicate that the motif is significant and non-significant.

Thus, the motifs are stored in a structured array with five members, which are listed besides the other used variables in table \ref{IO_variables}. Since the profile is just mirrored on the x-axis when hills are examined, the indices of the variables refer to dales.

The algorithm requires an equidistant sampled roughness profile, so the ordinate values of the profile $\bi{z}$ and the sampling interval $dx$ are needed as input. In addition, the feature type $FT$, the pruning type $PT$ and the threshold for pruning $TH$ must be specified as input for watershed segmentation. If segmentation according to dales is wanted, this is specified with $"\mathrm{D}"$ or $"\mathrm{V}"$. For hill with $"\mathrm{H}"$ or $"\mathrm{P}"$. $"\mathrm{V}"$ and $"\mathrm{P}"$ stand for pits and peaks and are relevant for the feature parameters which can specifically target peaks and pits (see section \ref{feature_parameters}). The pruning type $PT$ can be set to $"\mathrm{None}"$ if no pruning is to be performed. Otherwise, it can be specified which attribute should be pruned. $"\mathrm{Wolfprune}"$ for dale local depth, $"\mathrm{Width}"$ for dale local width, $"\mathrm{VolS}"$ for dale local volume and $"\mathrm{DevLength}"$ for dale local developed length. The corresponding threshold for pruning $TH$ can be freely selected or omitted if $PT$ is $"\mathrm{None}"$.
Watershed segmentation for profile sections can be divided into three steps, which are illustrated in figure \ref{steps}:
\begin{enumerate}
    \item determine indices of all peaks and pits (see section \ref{peakandpits})
    \item determine motifs (see section \ref{motifs})
    \item pruning (see section \ref{pruning})
\end{enumerate}

\subsection{Determine indices of all peaks and pits} \label{peakandpits}
The algorithm described in the following is summarized in the flowchart in figure \ref{step1}. First, the feature type being searched for must be considered. As already mentioned, when searching for hills or peaks, the profile is mirrored on the x-axis. The ordinate values of the profile are given by the vector $\bi{z}$, which therefore only needs to be multiplied by -1. As the algorithm progresses, it will always determine dale-motifs. This is possible because only the indices of the relevant points are stored. When the position indices are reinserted into the original (not mirrored) vector, the correct height information is obtained. Now to determine the local minima (pits) and local maxima (peaks), a vector is created listing the indices where neighboring height values are not equal, called $\bi{i_{\mathbf{Neq}}}$. Here a function is used that returns the indices of non-zero elements of a given vector in ascending order, which is used several times in the algorithm. The function is called \textbf{find}, like it is the case in Matlab. Similar functions or methods exist in other programming languages, such as Python (function numpy.where) or C\# (method FindIndex). However, there are syntax differences to consider. A flowchart of the \textbf{find} function used here can be found in the appendix (see \ref{find_sec}). If a second argument “First” is given to the function, only the first index is returned.
\begin{figure}
    \centering
    \input{figures/step1}
    \caption{Flowchart of step 1: Determination of all peaks and pits}
    \label{step1}
\end{figure}
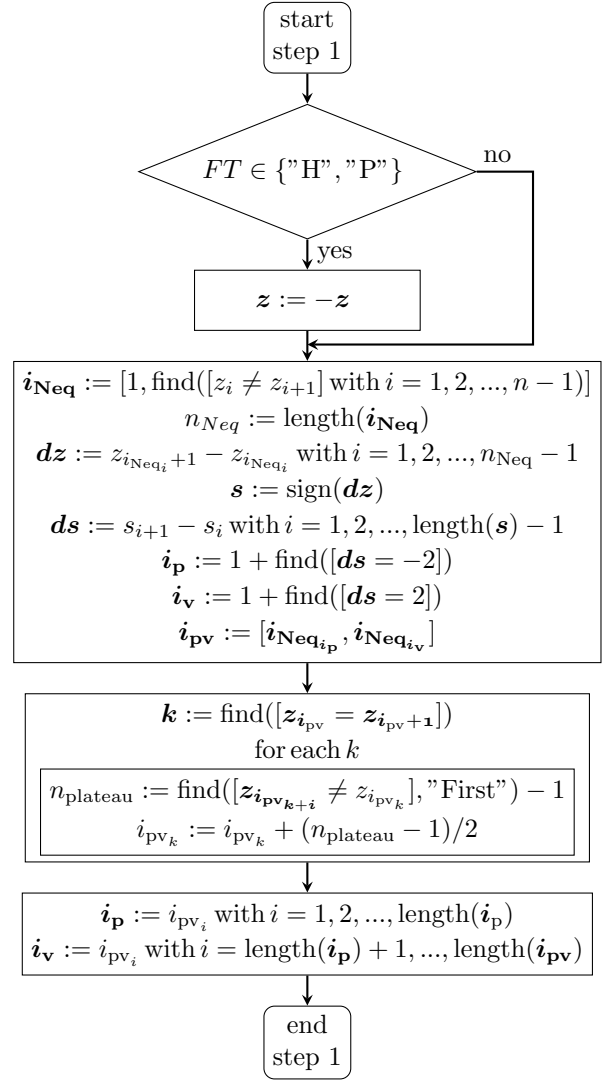
\begin{figure*}
    \centering
    \includegraphics{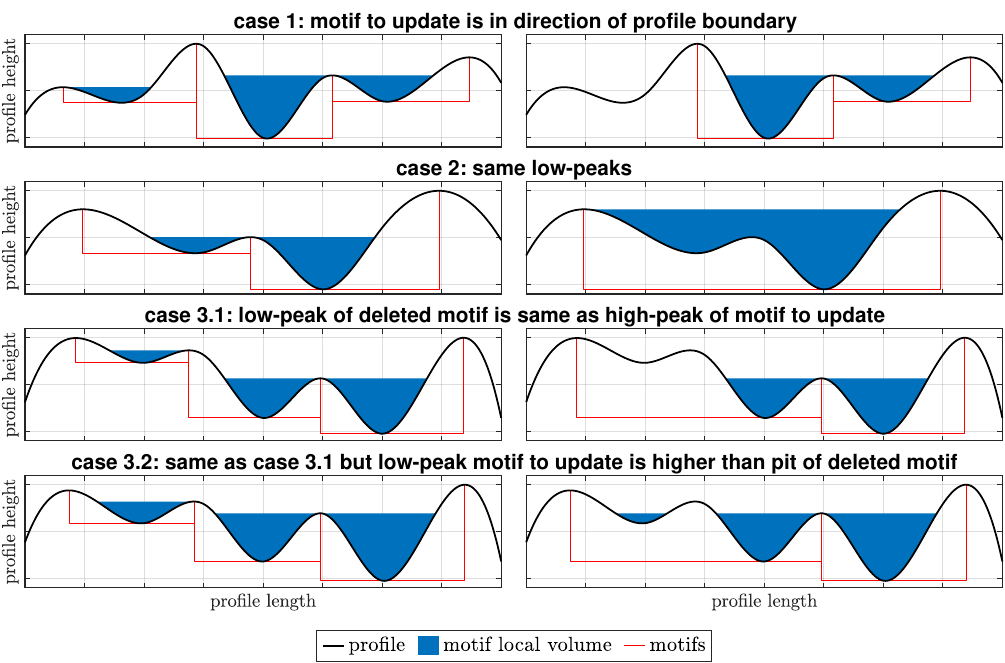}
    \caption{Three cases to distinguish in pruning. Left: unpruned motifs. Right: minimal motif pruned}
    \label{pruning_cases}
\end{figure*}
\begin{figure}
    \centering
    \input{figures/step2}
    \caption{Flowchart of step 2: Determination of all motifs}
    \label{step2}
\end{figure}
\begin{figure}
    \centering
    \input{figures/step3}
    \caption{Flowchart of the function prune\_min\_motif}
    \label{step3}
\end{figure}
For each ordinate value z, we want to determine whether the right neighbor is equal or not. For this task, an Iverson bracket with $[z_{i} \neq z_{i+1}]$ is used, which returns 0 if the statement is false or 1 if the statement is true. The resulting vector is passed to the \textbf{find} function, which returns the indices of the true statements.
Then, the differences between two consecutive elements of the vector $\bi{z}$ at the $\bi{i_{\mathbf{Neq}}}$ positions are calculated and the \textbf{signum} function is applied to it. The result is a sequence of $1$ and $-1$ that reflects whether the $z$-values are increasing or decreasing.
The indices of the local maxima and minima are then determined from the sequence $\bi{s}$. This identifies the positions in the sequence $\bi{s}$ where the algebraic sign changes from positive to negative (2) for maxima and from negative to positive (-2) for minima. Again, the combination of \textbf{find} function and Iverson bracket is used.
Next, the indices of the maxima and minima are transferred regarding the original index vector by inserting them into $\bi{i_{\mathbf{Neq}}}$ and the extrema are combined into a vector for the next section. This is because if the peak or pit is a plateau, the entry in $\bi{i_{\mathbf{pv}}}$ corresponds to the left edge of the plateau. However, the index of the center of the plateau needs to be stored. To achieve this, each entry is tested whether it is a plateau by checking if the right neighboring $z$-value is the same. If so, the number of values on this plateau $n_{\mathrm{plateau}}$ is determined by searching for the first $z$-value that is unequal again with the \textbf{find} function. The entry in $\bi{i_{\mathbf{pv}}}$ is then overwritten with the index plus half of $n_{\mathrm{plateau}}-1$. If $n_{\mathrm{plateau}}$ is odd, then $\bi{i_{\mathbf{pv}}}$ is a decimal number which is equal to the linearly interpolated index of peaks or pits. This step is completed by splitting $\bi{i_{\mathbf{pv}}}$ into the indices of the peaks $\bi{i_{\mathbf{p}}}$ and the indices of the pits $\bi{i_{\mathbf{v}}}$.
\subsection{Determine motifs} \label{motifs}
Now that the indices of the peaks and pits are known, the motif array ${M}$ can be filled as shown in figure \ref{step2}. A dale is always surrounded by two peaks. To ensure that only complete segments are considered, the pits to the left of the leftmost peak and to the right of the rightmost peak are removed from $\bi{i_{\mathbf{v}}}$. The total number of motifs $n_\mathrm{M}$ is then equal to the number of pits. For each pit, the corresponding interpolated indices of the low-peak, high-peak and the height-intersections are determined and entered into the motif array according to the convention (see section \ref{definitions}). All motifs are initially marked as significant (${{M}_k.sig=1}$). Since in the third step (pruning) the entries are updated if necessary, the steps for determining these values are delegated to the two functions \textbf{get\_ilp\_ihp} and \textbf{height\_intersections} (see \ref{get_ilp_ihp_sec} and \ref{height_intersections_sec}).

\subsection{Pruning} \label{pruning}
Initially, it is checked if pruning is desired by asking if the input pruning type is $"\mathrm{None}"$ (see figure \ref{step3}). If so, this step 3 is skipped. Otherwise, depending on the pruning type $PT$, the corresponding attribute values are determined for each motif. The function \textbf{feauture\_attribute} for determining the attribute values is outsourced because it is also needed for determining the feature parameters (see section \ref{feature_attribute_sec}). The attribute values are stored in a vector named $\bi{attr}$ with $n_\mathrm{M}$ entries. Now a while loop is started, which is running as long the minimal attribute value is lower than the threshold $TH$. First, the position of the minimum value in $\bi{attr}$ is determined, which also corresponds to the row in the motif array due to the same order and length. The corresponding motif is temporarily stored with ${M}_{\mathrm{min}}$, since some information may be needed later. Then the corresponding entries are deleted from the motif array and the $\bi{attr}$ vector.
The next step is to determine the row index of the motif to be updated ($r_\mathrm{U}$). This is the neighbor of the motif, into which the water would overflow if the deleted motif were further filled with water. I.e., from the pit towards the low-peak. This direction is stored in $dir$ as -1 for left or 1 for right. The row-index $r_\mathrm{U}$ is $r_\mathrm{min}-1$ if the motif to the left of the min-motif is to merge. If the right neighbor is to merge, then $r_\mathrm{U}$ is equal $r_{\mathrm{min}}$ because the min-motif is already deleted. The motif $r_\mathrm{U}$ may now need to be updated. There are three cases to distinguish:
\begin{enumerate}
\item The min-motif would overflow towards the border of the profile, where there is no motif. This is reflected in $r_\mathrm{U}=0$ or $r_\mathrm{U}>n_\mathrm{M}$ (see case 1 in figure \ref{pruning_cases}). If this is true, no further steps are needed.
\item The low-peak of the deleted motif and the motif to be merged are the same (see case 2 in figure \ref{pruning_cases}). Then the low-peak and the high peak of the motif to be merged are updated with the \textbf{get\_ilp\_ihp} function. The surrounding peaks are now the high-peaks of the motif to be merged and the deleted motif. The height intersection and attribute value of the motif to be merged are then updated.
\item The third case applies if the low-peak of the deleted motif is equal to the high-peak of the motif to be merged, or, more simply, if the first two cases do not apply. In this case, the high-peak of the motif to be merged is overwritten with the high-peak of the deleted motif. The further procedure depends on a condition and can be separated into the following two subcases: 
\begin{enumerate}
    \item If the height of the low-peak of the motif to be merged is lower or equal to the height of the pits of the deleted motif, no further steps have to be done (see case 3.1 in figure \ref{pruning_cases}).
    \item If the condition of 3.1 does not apply, then the pit of the deleted pit is below the water level of the new motif. Accordingly, the height intersection of the motif changes (see case 3.2 in figure \ref{pruning_cases}), which is updated in the next step. Except for the local height of the motif, the attribute value of the motif depends on the height intersection. Therefore, the attribute value is updated as well. An if-query was omitted here because the same steps are necessary for case 2, and calculating the local height does not add significant computational overhead.
\end{enumerate}
\end{enumerate}
\section{Feature parameters based on feature characterization} \label{feature_parameters}
The second subfunction of feature characterization derives feature parameters from the features obtained by the watershed segmentation. This requires the following three further sub-steps:
\begin{enumerate}
\setcounter{enumi}{3}
    \item determine significant features (see \ref{Significant_features_sec})
    \item determine the feature attributes of the significant features (see \ref{feature_attribute_sec})
    \item derive feature parameters by applying statistics to feature attribute values (see \ref{attribute_stats_sec})
\end{enumerate}
Therefore, the variables for describing the motif are required as input: $\bi{z}$, $dx$, ${M}$. Moreover, the subsequent variables are also necessary:\\[0.1cm]
\begin{tabular}{l p{0.78\columnwidth}}
$F_{\mathrm{sig}}$   & significant features $\in$ \{$"\mathrm{All}"$, $"\mathrm{Open}"$, $"\mathrm{Closed}"$, $"\mathrm{Top}"$, $"\mathrm{Bot}"$\}\\ 
$NI_{\mathrm{sig}}$  & nesting index for significant features $\in \mathbb{R}$\\
$AT$        & attribute type $\in$ \{$"\mathrm{HDh}"$, $"\mathrm{HDw}"$, $"\mathrm{HDv}"$, $"\mathrm{HDl}"$, $"\mathrm{PVh}"$, $"\mathrm{Curvature}"$, $"\mathrm{Count}"$\}\\
$A_\mathrm{stats}$ & attribute statistics $\in$ \{$"\mathrm{Mean}"$, $"\mathrm{Max}"$, $"\mathrm{Min}"$, $"\mathrm{StdDev}"$, $"\mathrm{Perc}"$, $"\mathrm{Hist}"$, $"\mathrm{Sum}"$, $"\mathrm{Density}"$ \} \\
$v_\mathrm{stats}$ & required value if $A_\mathrm{stats}$=$"\mathrm{Perc}"$ $\in \mathbb{R}$
\end{tabular}\\

\subsection{Significant features} \label{Significant_features_sec}
\begin{figure*}
    \centering
    \includegraphics{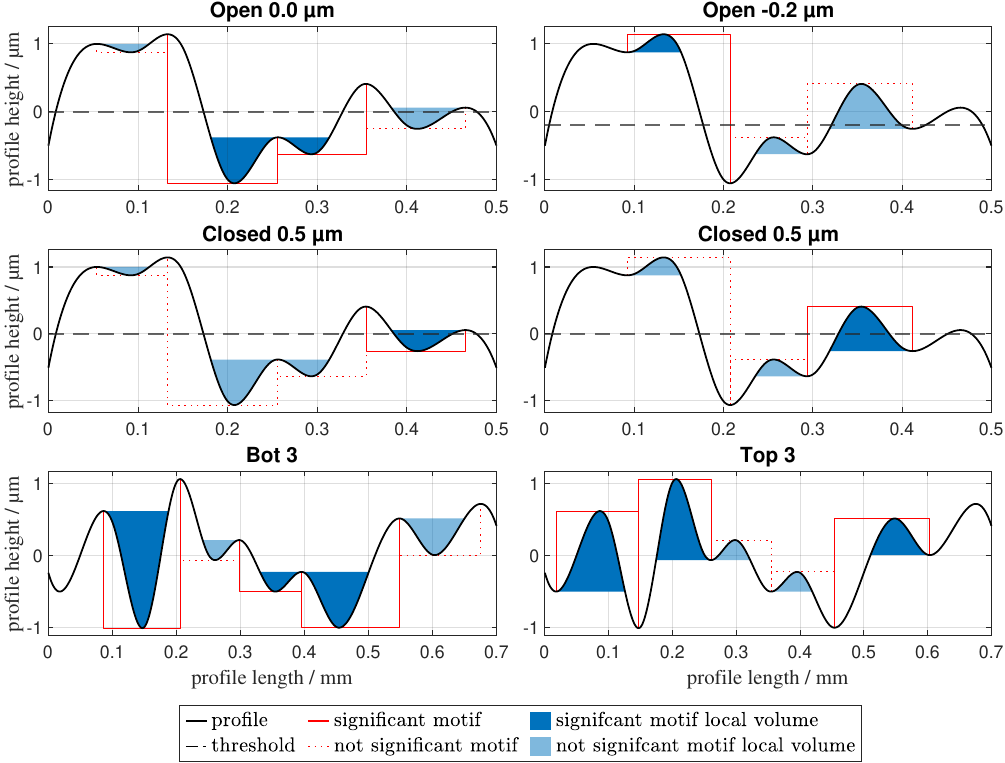}
    \caption{Sample profiles for the three different concepts of selecting significant features}
    \label{step4_visu}
\end{figure*}

A distinction can now be made between significant and non-significant features, whereby the significant features are used for the further calculation of the feature parameters. According to ISO 21920-2 \cite{21920-2}, a distinction is made between three concepts for the selection of significant characteristics, which have been derived from ISO 25178-2 \cite{25178-2}. However, the list of methods is still open for expansion. The algorithm to achieve the distinction is illustrated in figure \ref{step4}, while corresponding examples are shown in Figure \ref{step4_visu}.
\begin{figure}
    \centering
    \input{figures/step4}
    \caption{Flowchart of step 4: Determine significant features}
    \label{step4}
\end{figure}

\begin{enumerate}
    \item $"\mathrm{All}"$: All features are used for further determination.
    \item $"\mathrm{Open}"$ and $"\mathrm{Closed}"$: The concept of open and closed motifs is based on a mechanical-rheological model by Sobis \cite{SOBIS1992233}, was further developed from Weidel \cite{Weidel.2007} and was transferred to ISO25178-2 \cite{25178-2} by Blateyron \cite{Blateyron.2016} as a correction in 2021. In this model, a height threshold is introduced to simulate the plastic deformation depth that occurs when a load is applied to the surface. Dales are delineated from the top, and the fluid they contain may or may not escape depending on the threshold. If the threshold is higher than the lowest saddle point, there is still space above the ridge line from which the liquid can escape. In this case, the dale is considered open. If the sill is lower than the lowest saddle point, the dale is surrounded by a wall set at height c and nothing can escape. A practical example of this method applied to steel sheets can be found in \cite{BATALHA2001732}. To be able to transfer the model to the profile section, it must be assumed that the water can only move in the plane of the profile section. Otherwise, there is always the possibility that a dale is open to another dale in the direction of the third dimension. Consequently, a dale is considered closed if the threshold is below the lower peak and above the pit. A dale is deemed open if the threshold height is above the lower pit. 
    In the Annex of ISO 21920-2 \cite{21920-2} a dale is closed if the threshold height is below the lower pit, which means that a dale is also be considered closed if the threshold height is lower than the pit. According to the model, there would no longer be a dale due to the plastic deformation and therefore no enclosed fluid. In ISO 25178 \cite{25178-2} there is the addition that fluid from within the dale cannot get into a neighboring dale. We recommend adding this also to ISO-21920-2 \cite{21920-2}.
    \item $"\mathrm{Top}"$ and $"\mathrm{Bot}"$: According to ISO 21920-2 \cite{21920-2}, with “Bot” the pits can be declared significant that belong to the top $N$ pit depths. With “Top” the peaks that belong to the top $N$ peak heights are significant. This assumes that either peaks ($FT="\mathrm{P}"$) or pits ($FT="\mathrm{V}"$) have been segmented. In this implementation, however, it is also possible to segment by hills or dales. In this case, the dales are significant, whose pits belong to the top $N$ pit depths.
\end{enumerate}
Since all features have been declared as significant so far, the non-significant features are now determined and their indicator ${M}.sig$ is set to zero. To achieve this, the variable $\bi{I_\mathbf{Nsig}}$ is introduced, which is filled with the indices of the non-significant features and is initialized as an empty vector.
\begin{figure*}
    \centering
    \input{figures/func_feature_attribute}
    \caption{Flowchart of the function feature\_attribute respectively step 5}
    \label{step5}
\end{figure*}
If “All” is requested, nothing further needs to be done, as all features are already declared as significant.

If “Top” or “Bot”, $NI_{\mathrm{sig}}$ is first overwritten with $n_\mathrm{M}$ if $n_\mathrm{M}$ is smaller. The peak heights or pit depths are then determined and stored in $\bi{h}$. To determine the top $N$ from this, $\bi{h}$ is sorted in descending order and thus the indices are determined from the largest $\bi{h}$ ($i_1$) to the smallest $\bi{h}$ ($i_{n_\mathrm{M}}$). The indices of the features whose pits or peaks do not belong to the top $NI_{\mathrm{sig}}$ are then stored in the vector $\bi{I_\mathbf{Nsig}}$. ${M}.sig$ is then set to 0 for these indices.
If “Open” or “Closed” is requested, the heights of low-peaks or high-peaks are determined first. Then a feature type indicator is determined by calculating the difference between the height of the low-peak and the pit and applying the \textbf{signum} function.

If “Open”, the indices motifs are determined for which the low-peak is greater than the threshold value $NI_{\mathrm{sig}}$, i.e., are not significant. Both $z_{\mathrm{lp}}$ and $NI_{\mathrm{sig}}$ are multiplied by $FTI$. In the case of dales ($FTI=1$), nothing changes in the expression. However, in the case of hills ($FTI = -1$), the larger turns into smaller. If “Closed” features are sought, the heights of the pits or peaks $z_\mathrm{v}$ are determined first. The indices of the non-significant features are then determined by examining which feature $z_{\mathrm{lp}}$ is smaller than $NI_{\mathrm{sig}}$ or $z_\mathrm{v}$ is larger than $NI_{\mathrm{sig}}$. $FTI$ also ensures here that the inequalities are reversed if the features are hills. The step is completed by using $\bi{I_\mathbf{Nsig}}$ to set all non-significant features ${M}.sig=0$.

Depending on the configuration, errors may occur in this step. On the one hand, if the pruning threshold is set too high, the motif array is empty. On the other hand, errors occur if all features have been declared as not significant. The handling of these errors has been omitted from the flowchart for the sake of traceability. In the Matlab implementation in the appendix (see section \ref{feature_parameter_appendix}), both errors are intercepted by an if-query. If an error is encountered, a corresponding warning is issued. The values of $\bi{x_{\mathbf{FC}}}$ and $\bi{attr}$ are set to $\mathrm{NaN}$.
\subsection{Feature attributes} \label{feature_attribute_sec}
As mentioned above, the determination of the feature attributes is already necessary for pruning (step 3) in the watershed segmentation. This step was therefore embedded in a function, which is shown in figure \ref{step5}, to determine the attribute values of the significant motifs. In addition to $\bi{z}$ and $dx$, the motif array and the attribute type or pruning type are required as input. When pruning, it is sometimes necessary to update attribute values for individual motifs. In this case, it is sufficient to specify the corresponding line of the motif array as input.
First, the \textbf{find} function is used to determine the indices of the significant motifs for which the attribute values sought are then determined in the next step.
For these significant motifs, the attribute values are then determined according to the given attribute type $AT$ or pruning type $PT$ and stored in a vector $\bi{attr}$, which is returned as the output of the function. In the case of a plateau, the indices may not be integers, so they are floored, which by definition does not change the height value. The attributes are determined as follows:
\begin{itemize}
\item "HDh” or “Wolfprune” for dale local depth:\\ The absolute value of the height difference between low-peak and the pit is calculated. 
\item “HDw” or “Width” for dale local width:\\ The amount of the difference between the lateral position of the low-peak and height intersection is formed. Due to the equidistant sampled profile, it is sufficient to calculate the difference between the indices and multiply it by $dx$. As there may be several height intersections, the maximum value of the absolute values is determined.
\item "PVh" for pit depth:\\ The feature type indicator introduced in step 4 is determined here, which allows the sign to be reversed if hills are involved. 
\item "Count" for Count:\\ The value 1 is assigned to each attribute value, which later allows the density of the (significant) features to be determined.
\end{itemize}
The following determinations of attribute values were assigned to functions described in the appendix, as they are much more extensive:
\begin{itemize}
\item "VolS or "HDv" for dale/hill local volume (see \ref{func_HDvf_sec})
\item "DevLength" or "HDl" for Dale/Hill local length (see \ref{func_HDlf_sec})
\item "Curvature" for Curvature (see \ref{func_Curvature_sec})
\end{itemize}

\subsection{Attribute statistics} \label{attribute_stats_sec}
\begin{figure*}
    \centering
    \includegraphics{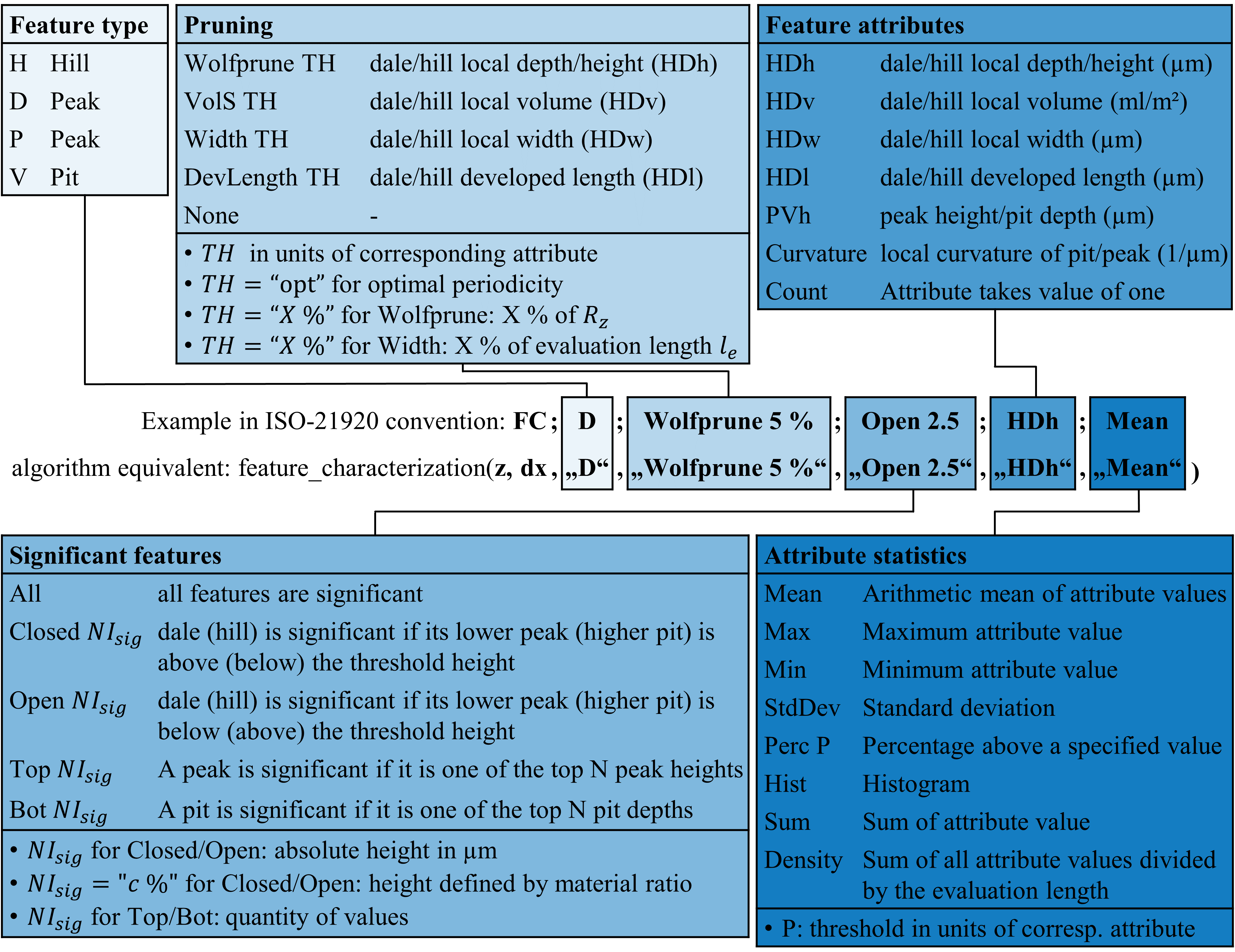}
    \caption{Illustration of the feature characterization toolbox. It demonstrates how the individual tools are specified in accordance with the ISO 21920 convention and which potential tools can be employed. Furthermore, it illustrates how this example is adapted with the function feature\_characterization.}
    \label{FC_Convention}
\end{figure*}

The last step 6 is to derive a feature parameter from the attribute values of the significant features using statistics or alternatively a histogram of the attribute values. The corresponding algorithm is visualized in figure \ref{step6}. The input variable $stats$ specifies which statistical value is to be determined. The selection options with description are shown in figure \ref{FC_Convention}. Another input called $v_\mathrm{stats}$ is required for $Perc$, which specifies the limit value for the percentage of attribute values above it.
\begin{figure}
    \centering
    \input{figures/step6}
    \caption{Flowchart of step 6: Attribute statistics}
    \label{step6}
\end{figure}
\section{Feature characterization} \label{FC_sec}
With the main function \textbf{feature\_characterization}, the \textbf{watershed\_segmentation} and \textbf{feature\_para-meter} function can be executed in one step. The input arguments of the function are based on the convention of the notation described in ISO-21920-2 (Annex F.7). The various tools are listed in a sequence and separated by “;”. The same designations for the tools can be specified in the same order as input arguments of the function as strings. Only $\bi{z}$ and $dx$ have to be prefixed. Figure \ref{FC_Convention} shows an example in which the tools that can be used are summarized once again in the connected boxes. It is anticipated that further selection options will be added in future revisions of ISO 21920-2 \cite{21920-2}.
The function \textbf{feature\_characterization} primarily serves as a parser that decomposes the inputs and, if necessary, converts them into suitable formats for the two subfunctions \textbf{watershed\_segmentation} and \textbf{feature\_parameter}. A detailed description of the function can be found in \ref{func_FC_sec}.

 \subsection{Named parameters}
The feature parameters named in ISO 21920-2 \cite{21920-2} are listed in table \ref{tab:feature_parameter}. These correspond to the named areal feature parameters from ISO 25178-2 \cite{25178-2}. 

\section{Example profiles and potential applications} \label{example_profiles}
In this section the presented algorithm is applied to both artificial and practical profiles to demonstrate its plausibility and potential applications. A thorough investigation of the algorithm's stability and robustness should be conducted in future work using diverse test profiles. Blateyron previously demonstrated in \cite{Blateyron_2021} that feature characterization is robust against factors such as lateral shift, direction, and noise. However, this was based on the FDIS version of ISO 21920-2, without revealing the algorithm used. Therefore, future work should verify these results using the presented algorithm.

\begin{figure*}
    \centering
    \includegraphics{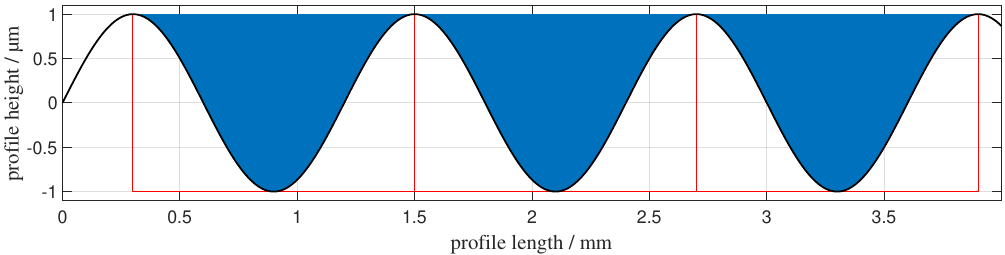}
    \caption{Evaluation of a sinusoidal profile with wavelength $1.2\, \mathrm{mm}$}
    \label{case_study1}
\end{figure*}

In figure \ref{case_study1} is a sinusoidal profile with an amplitude of 1 µm and a wavelength of $\SI{1200}{\micro\meter}$. The step size was set to $dx=\SI{0.5}{\micro\meter}$. The profile has been segmented according to dales. Pruning is not necessary here because there is no noise. The table \ref{tab:feature_parameter} lists several feature parameters. For comparison, the theoretical value for the continuous sine has been calculated and the deviation from each feature parameter is shown in table \ref{tab:feature_parameter}.
\begin{table*}
    \caption{Named feature parameters according to ISO 21920-2 and their equivalent in ISO 25178-2}
    \centering
    \begin{tabular}{|p{3.41cm}|p{6.79cm}|p{3.19cm}|p{2.01cm}|}
    \hline
    Description& FC Convention & Feature parameter & Equivalent in \\
     &  & in ISO 21920-2 &  ISO 25178-2 \\
    \hline
    Density of peaks                & FC; P; Wolfprune 5 \%; All; Count; Density    & Ppd, Wpd, Rpd     & Spd\\
    Density of pits                 & FC; P; Wolfprune 5 \%; All; Count; Density    & Pvd, Wvd, Rvd     & Svd\\
    Mean peak curvature             & FC; P; Wolfprune 5 \%; All; Curvature; Mean   & Pmpc, Wmpc, Rmpc  & Spc\\
    Mean pit curvature              & FC; V; Wolfprune 5 \%; All; Curvature; Mean   & Pmvc, Wmvc, Rmvc  & Svc\\
    Five-point peak height          & FC; P; Wolfprune 5 \%; Top 5; PVh; Mean       & P5p, W5p, R5p     & S5p\\
    Five-point pit depth            & FC; V; Wolfprune 5 \%; Top 5; PVh; Mean       & P5v, W5v, R5v     & S5v\\
    Ten-point height                & R5p+R5v                                       & P10z, W10z, R10z  & S10z\\
    \hline
    \end{tabular}
    \label{tab:named_parameters}
\end{table*}
\begin{table*}
    \caption{Different feature parameter for sinusoidal profile in figure \ref{case_study1}}
    \centering
    \begin{tabular}{|p{5.5cm}|p{5cm}|p{2.9cm}|p{2cm}|}
    \hline
    feature parameter & FC Convention & result & deviation to theo. value \\
    \hline
     Mean dale local height             & FC;D;None;All;HDh;Mean        & 2 µm                                  & 0 \% \\
     Mean dale local width              & FC;D;None;All;HDw;Mean        & 1200 µm                               & 0 \% \\
     Mean dale local volume             & FC;D;None;All;HDv;Mean        & 0.3 ml/m²                             & 3.7e-14 \%\\
     Mean dale local developed length   & FC;D;None;All;HDl;Mean        & 1200.008224 µm                        & 3.9e-10 \%\\
     Mean pit curvature                 & FC;D;None;All;Curvature;Mean  & 2.741556e-5 µm\textsuperscript{-1}    & 3.33e-9 \%\\
    \hline
    \end{tabular}
    \label{tab:feature_parameter}
\end{table*}
\begin{figure*}
    \centering
    \includegraphics{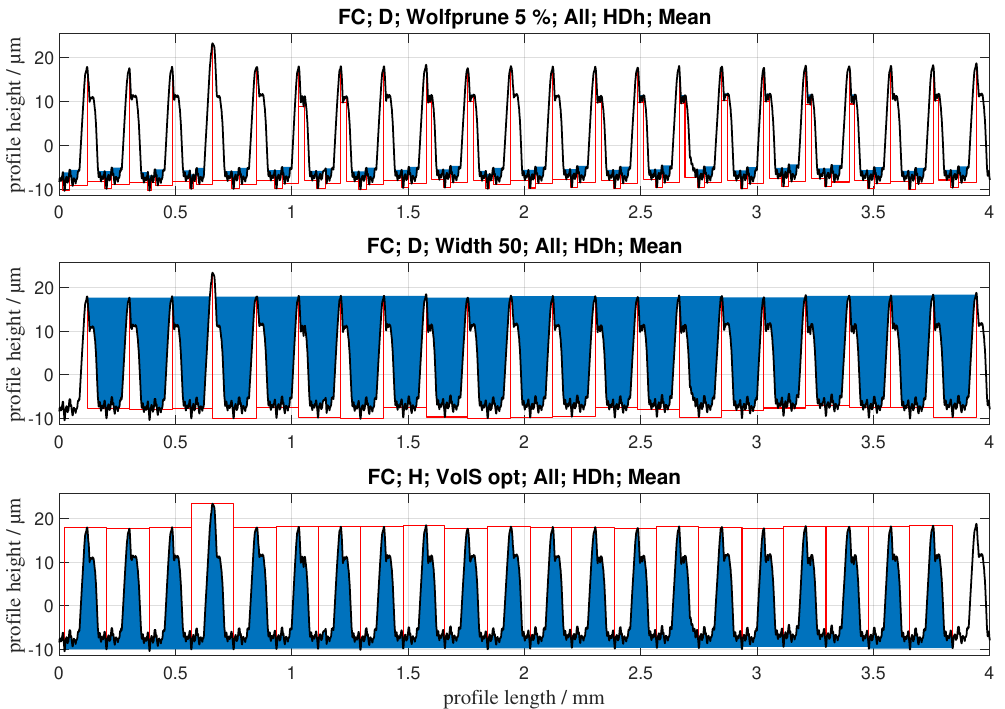}
    \caption{Profile of a turned surface with three different configurations of watershed segmentation. The applied configuration of the watershed segmentation can be found in the title of the respective plot.}
    \label{case_study_turning}
\end{figure*}

Figure \ref{case_study_turning} shows a profile of a turned surface with three different configurations for pruning. In the upper profile, the default value according to ISO 21920-3 \cite{21920-3} for Wolf pruning was used, which is $\SI{5}{\percent}$ of $Rz$. In instances where the objective is to extract the features that were cut into the material by the tool's cutting edge, this would be classified as an instance of oversegmentation. On the other hand, this setting can also be purposeful when the aim is to recognize and quantify such smaller features. For instance, they could provide insights into the state of the manufacturing process, e.g., tool wear. Moreover, critical points could be identified that might potentially lead to crack formation and component failure. No pruning would be equally sensible in this case, provided that any potential noise has already been eliminated through filtering. In the case of a deterministic surface, such as the one in question, it is typical to seek features that are as consistent as possible. In this instance, this would mean identifying those produced by the tool's cutting edge, as seen in the middle profile. Due to the process, the distance between these features corresponds to the feed rate. Therefore, it is logical to prune based on width and select a percentage of the feed rate as the threshold. If the feed rate is unknown, the average distance between the profile elements, $RSm$, could be employed. Nevertheless, to provide recommendations for thresholds for the different pruning types, which may also vary depending on the application, a comprehensive investigation is necessary. This paper can serve as a foundation for such an investigation. This example clearly illustrates that the pits of the dales do not necessarily have to correspond to the lowest points within the dales. The same applies to the peaks of the hills. This is because these pits of the dales are deleted during pruning due to insufficient width/volume/dev. length. Only in Wolf pruning does the pit of the dale always correspond to the lowest point within the dale. That means even if the motifs regarding low and high peaks are the same, different pruning methods can lead to significant deviations when calculating curvature, dale/hill local height (HDh), or peak height/pit depth (PVh).
\begin{figure*}
    \centering
    \includegraphics{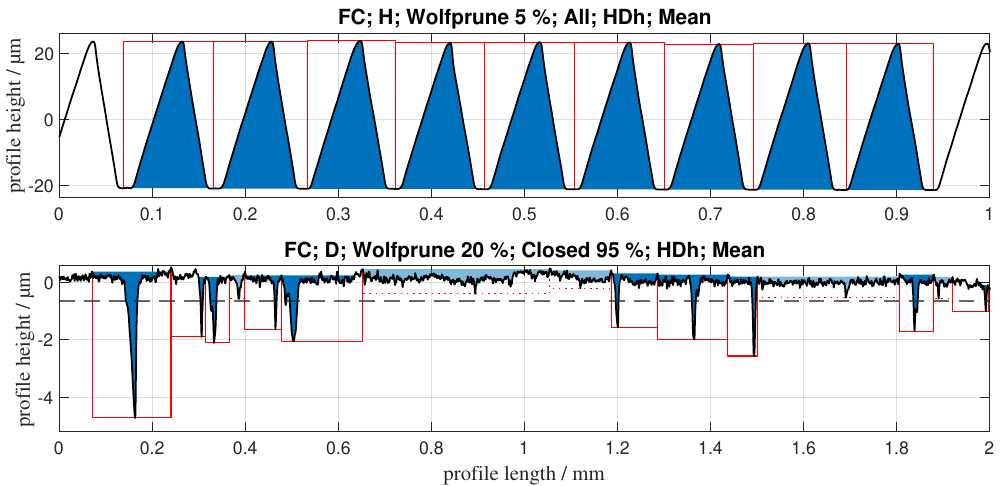}
    \caption{Profile of a riblet surface (top) and profile of a honed surface (bottom). The applied configuration of the watershed segmentation can be found in the title of the respective plot.}
    \label{case_study_function}
\end{figure*}
\begin{figure*}
    \includegraphics[width=1\textwidth]{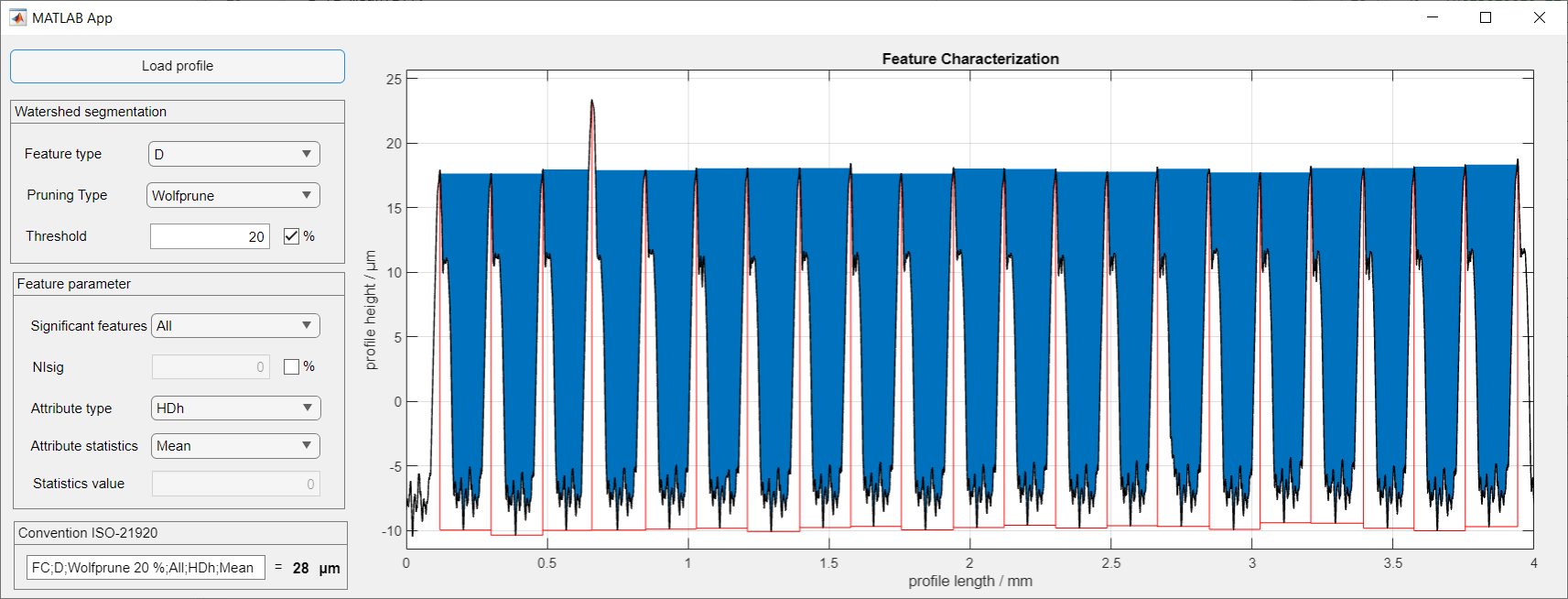}
    \caption{Graphical user interface}
    \label{GUI}
\end{figure*}
There is usually particular interest in the peaks of the profile, as they are associated with the function (e.g., friction, wear). Therefore, the lower profile is segmented after peaks as an example. Unlike the dales, the width of the hills is smaller due to distribution and is no longer almost equal to the feed rate, which must be considered in the threshold.
As an alternative approach, the \textbf{optimal\_periodicity} function was employed (see \ref{func_optimal_TH_sec}), which is not part of ISO 21920-2 \cite{21920-2}. This function determines the threshold at which the motifs are as similar as possible, thereby identifying the optimal periodicity. 
For example, $Rpd$, density of peaks, and $Rmpc$, mean curvature of peaks, can be determined, which are the equivalents to the areal feature parameters $Spd$ and $Spc$. These are usually related to the function of the surface (see section \ref{introduction}).

Another example of a deterministic surface is shown in figure \ref{case_study_function} at the top. These are riblets that are intended to reduce the flow resistance of the component. Feature characterization allows the riblets to be quantified. By extending the functionality, further relevant parameters could be calculated, such as the average distance between the peaks or a measure of the deviation from the target geometry.

In figure \ref{case_study_function} at the bottom shows a profile of a honed surface.
In the majority of honed profile cases, the established $Rk$ parameters would be employed, which are derived from the Abott Firestone curve. These parameters enable the separate evaluation of the core, peak, and valley area. Each of these areas holds different functional implications.
For instance, the area of dales $Rak2$ (formerly  $A2$) is a measure for the protruding pits and thus the oil retention volume of the surface. Woś was able to show in \cite{Wos.2011} that this parameter correlates negatively with friction. However, this correlation cannot be universally extended to other applications with honed surfaces \cite{Mezghani.2013,Yousfi_2014}.
With feature characterization, the oil retention volume of the deep grooves could be calculated more specifically. Furthermore, a statement could be made regarding the density or distribution of the deep grooves, which is likely to influence the tribological behavior when lubricant is used. This can be achieved by considering only the closed motifs. The limit value must be selected so that the dales with pits at the height of the plateau are declared as not significant. In this case, Closed 95\% was selected, i.e., the inverse material ratio at 95\% ($Rcm$(95\%)), as defined in ISO 21920-2 \cite{21920-2}.

\section{Availability and softgauge} \label{implementation}
The described algorithm has been implemented in Matlab (see appendix \ref{Matlab}) and Python. Both implementations can be found on GitHub\footnote{https://github.com/mts-public/feature-characterization-for-profile-surface-texture}. In addition to the code, we included a minimal example to test the functionality for the first time. The profiles shown and others are stored in the data folder. There is also a graphical user interface (GUI) that can be run as a standalone after installation (see figure \ref{GUI}). In the GUI, the numerous options can be selected via drop-down menus and the thresholds can be changed via an input field. The effects on the segmentation and the result are visible after each change, which makes it easier to get started and to determine the appropriate settings, especially for pruning. 

The provided implementation can be used as reference software (softgauge of type F2 according to ISO 5436-2 \cite{5436-2}). For this purpose, there is a script in the “Softgauge” folder with which reference data (softgauges of type F1) can be loaded. The named parameters (see table \ref{tab:named_parameters}) are calculated, which serve as a reference for your implementation. The reference data must have the structure and format described in ISO 5436-2 \cite{5436-2} (.smd). Such profiles can be obtained from npl.com, for example.

It is planned that both the GUI and the softgauge application will be accessible via a web-based interface. A link will be provided in the GitHub repository.
\section{Conclusion}
Feature characterization for profile surface texture as standardized in ISO 21920-2 \cite{21920-2} is a versatile tool for extracting and quantifying surface features relevant to various surface properties. In this paper we present an unambiguous algorithm in pseudocode, that has been implemented in both Matlab and Python and is publicly available on GitHub (see section \ref{implementation}). The modular design of the algorithm allows the combination of all relevant process steps of the standardization, demonstrated through practical examples that confirm the plausibility of the code (see section \ref{example_profiles}).
In future work, the existing default value for Wolf pruning should be investigated in more detail. In addition, recommendations for threshold values for pruning by VolS, DevLength and Width should be developed, which could possibly be included in future revisions of ISO 21920-3 \cite{21920-3}.
In addition, there could also be an exploration of how the toolbox could be extended. For example, to: 
\begin{itemize}
 \item assess the shape (e.g., roundness) of the features equivalent to ISO 25178-2 \cite{25178-2},
 \item declare features as significant if they are above or below a specified threshold, or
 \item assess the lateral or vertical distribution of features.
\end{itemize}
Especially for anisotropic or semi-anisotropic surfaces, such as those created by turning, grinding, or microstructuring processes, feature characterization should be investigated in detail. Here, the feature parameters based on feature characterization could provide further opportunities to assess the function of the surface and the properties of the manufacturing process.
\ack
This work was funded by the Deutsche Forschungsgemeinschaft (DFG, German Research Foundation)—
project number 461839204.
\clearpage

\section*{References}

\input{main.bbl}
\clearpage
\appendix
\section{}
\subsection{Function find} \label{find_sec}
The find function scans an array $\bi{A}$ and returns the indices of those elements that are not zero. Optionally, the function can return only the first index of such an element if the $FirstOnly$ parameter is set accordingly. If this parameter is not passed, the default is 0 and the function returns all relevant indices. The function starts with initialization, during which it checks if $FirstOnly$ is set. It then iterates over each element of the array $\bi{A}$. For each element, it checks if it is not equal to zero. If this condition is met, the corresponding index is included in the result array $\bi{i_{\mathbf{N0}}}$. If $FirstOnly$ is 1, the search for the first non-zero element is aborted and the corresponding index is returned immediately. Finally, the function returns the array $\bi{i_{\mathbf{N0}}}$ containing either all relevant indices or only the first index, depending on the configuration of the $FirstOnly$ parameter.
\subsection{Function get\_ilp\_ihp} \label{get_ilp_ihp_sec}
This function takes the vector $\bi{z}$ and the indices of the two surrounding peaks $\bi{i_{\mathbf{p}_{\mathbf{surr}}}}$ of the examined pit as input. The $z$-values of the indices $\bi{i_{\mathbf{p}_{\mathbf{surr}}}}$ are stored in a vector $\bi{z}_{\mathbf{surr}}$ and then the index within $\bi{z}_{\mathbf{surr}}$ of the lower peak is determined (see figure \ref{get_ilp_ihp}). If both $\bi{z}$ values are the same, the first value is assumed to be lower. Accordingly, the index of the lower peak $i_{\mathrm{lp}}$ and the index of the higher peak $i_{\mathrm{hp}}$ are returned.
\begin{figure}
    \centering
    \input{figures/func_find}
    \caption{Flowchart of the function find}
    \label{find}
\end{figure}
\begin{figure}
    \centering
    \input{figures/func_get_ilp_ihp}
    \caption{Flowchart of the function get\_ilp\_ihp}
    \label{get_ilp_ihp}
\end{figure}

\subsection{Function height\_intersections} \label{height_intersections_sec}
This function is based on the crossing-the-line algorithm by Seewig et al. \cite{Seewig.2020}, where the first step is to search for intersections with the reference line. However, the “region of doubt”, which should take noise into account, has not been adopted, since noise is considered here by pruning. Figure \ref{height_intersections} shows the flowchart of the function. The z-profile, the index of low-peak, and high-peak are required as input. 
First, an empty vector $\bi{i_{\mathbf{hi}}}$ is created, which is subsequently filled with the height intersections found. Then the direction in which the height intersection is to be searched for is determined. To achieve this, the difference between $i_{\mathrm{lp}}$ and $i_{\mathrm{hp}}$ is entered into the \textbf{signum} function. The result $dir=-1$ means left and $dir=1$ means right. The two indices $i_{\mathrm{lp}}$ and $i_{\mathrm{hp}}$ are rounded, as they may be interpolated indices. Then the height value of the low-peak $z_{\mathrm{lp}}$ is determined. The initialization is completed by defining the starting point $j$, which is equal to $i_{\mathrm{lp}}$ when the lower peak is not a plateau. In the case of a lower peak that is a plateau, however, the index from the edge of the plateau towards the pit must be used, as otherwise this edge will already be recognized as a height intersection. This is achieved by starting from index $i_\mathrm{lp}$ and searching for the first index whose height value is not equal to $z_\mathrm{lp}$. From this index now one is subtracted, the result is multiplied by $dir$ and added to $i_\mathrm{lp}$ to determine the index of the last plateau value.
Searches are only carried out up to the index of the higher peak. The loop checks for each index within the specified range whether the height $z_{\mathrm{lp}}$ is intersected within an increment. If so, the position of the height intersection is determined by linear interpolation and appended to the vector $\bi{i_{\mathbf{hi}}}$.

\begin{figure}
    \centering
    \input{figures/func_height_intersections}
    \caption{Flowchart of the function height\_intersections}
    \label{height_intersections}
\end{figure}

\subsection{Function HDvf} \label{func_HDvf_sec}
\begin{figure}
    \centering
    \input{figures/func_HDvf}
    \caption{Flowchart of the function HDvf}
    \label{func_HDvf}
\end{figure}
The steps to determine the dale local volume are illustrated in figure \ref{func_HDvf}. $\bi{z}$, $dx$ and the respective motif ${M}_\mathrm{r}$ are required as input of the function. During initialization, the height of the low-peak $z_{\mathrm{lp}}$ is determined first. The direction $dir$ is then determined, indicating the location of the high-peak in relation to the low-peak ($dir=1$ for right, $dir=-1$ for left). This information is necessary for determining the counting direction and for ceil or floor for interpolated indices. Pruning may result in multiple intersections within the motif, meaning that the area may consist of several partial areas. The vector $\bi{i_{\mathbf{hi}}}$ is used to assign the intersections, starting with $i_{\mathrm{lp}}$ as the first value, followed by the height intersections from ${M}_\mathrm{r}$. The height intersections are always sorted from the low-peak to the high-peak due to the method of determination of height intersections (see \ref{height_intersections_sec}).

In the subsequent loop, the areas are calculated and added up using the vector $\bi{i_{\mathbf{hi}}}$ from odd indices to the next index ($i_{\mathrm{hi}_1}$ to $i_{\mathrm{hi}_2}$, $i_{\mathrm{hi}_3}$ to $i_{\mathrm{hi}_4}$, etc.). To accomplish this, the initial values of $A=0$ and $i=1$ are set. In this section, vectors $\bi{x}_\mathbf{f}$ and $\bi{z}_\mathbf{f}$ are created to describe the path between the height intersections. To achieve this, integer indices are required within the respective area. Interpolated indices are rounded towards the center of the respective partial area. If the counting direction is to the right ($dir=1$), the first value must be rounded up and the second rounded down. However, if the counting direction is to the left ($dir=-1$), rounding is reversed. To do this, the indices are multiplied by the value of $dir$, then rounded, and the absolute value is taken. If $dir=1$, this does not affect the result. However, if $dir=-1$, the amount is rounded in the opposite direction. The negative sign is then removed by taking the absolute value.
The vectors $\bi{x_\mathbf{f}}$ and $\bi{z_\mathbf{f}}$ can now be generated with these integer indices $i_1$ and $i_2$. As the profile is equidistant sampled, it is sufficient to increment or decrement $\bi{x_\mathbf{f}}$ from $i_1$ to $i_2$ and multiply by $dx$. For $\bi{z_\mathbf{f}}$, the corresponding $z$-values are determined from $i_1$ to $i_2$.
To describe the complete path, the unrounded height intersections ${i_{{\mathrm{hi}}_i}}$ and ${i_{{\mathrm{hi}}_{i+1}}}$ are added in the case of $\bi{x_\mathbf{f}}$. In the case of $\bi{z_\mathbf{f}}$ with the corresponding value $z_\mathrm{lp}$ as defined. If ${i_{{\mathrm{hi}}_i}}$ and ${i_{{\mathrm{hi}}_{i+1}}}$ are integers, there will be duplications at the beginning and end of $\bi{x_\mathbf{f}}$. However, since there is no distance between the two entries, there is no area and therefore no influence on the result. The area can now be calculated using the trapezoidal rule. Most programming languages have this embedded as a function (Matlab: trapz, python: numpy.trapz). The absolute value of this is added to the value of the variable A.
The loop is completed by adding 2 to the running variable i so that the next partial area is calculated if necessary. The result of $A$ is divided by the evaluation length and returned as the output $HDv$ of the function. According to ISO 21920-2 \cite{21920-2}, the result must be given in $\mathrm{\frac{ml}{m^2}}$. If both z and $dx$ are given in \unit{\micro\meter}, no conversion is required.

\subsection{Function HDlf} \label{func_HDlf_sec}
The \textbf{HDlf} function is illustrated in figure \ref{func_HDlf}. The function requires $\bi{z}$, $dx$, and the motif ${M}_\mathrm{r}$ to be examined as input. As well as the calculation of $z_\mathrm{lp}$ and $dir$ first. Only the last height intersection starting from the low-peak $i_{{\mathrm{hi}}_{end}}$ is necessary. This corresponds to the last entry in ${M}_\mathrm{r}.\bi{i_{\mathbf{hi}}}$. The integer indices that limit the values are determined by rounding in the direction of the pit, following the same scheme as in the function \textbf{HDvf}. The resulting $z$-values are stored in vector $\bi{z}_\mathbf{f}$, which has integer indices. A difference vector, $\bi{dz}$, is then calculated to represent the vertical distances between neighboring values. This information is used to calculate the developed length, which consists of three parts. The initial section consists of values with integer indices, whose vertical distances have been determined using $dz$. The horizontal distances between these points are always $dx$, due to the equidistant profile. The linear distance between points can now be calculated and added up using Pythagoras' theorem. The second section is only relevant if the low peak is a plateau that has a non-integer value. In this case, the modulo of ${M}_\mathrm{r}.i_\mathrm{lp}$ is taken and multiplied by $dx$. By definition, there is no difference in height here. If ${M}_\mathrm{r}.i_\mathrm{lp}$ is an integer, this part equals 0. The third part is the distance between the last integer index $i_2$ and the last height intersection $i_{{\mathrm{hi}}_{end}}$. Here, we calculate the horizontal and vertical distance between the last integer index $i_2$ and the height intersection $i_{{\mathrm{hi}}_{end}}$ using the Pythagorean theorem. If $i_{{\mathrm{hi}}_{end}}$ is already an integer, both the horizontal and vertical distances are equal to 0.
\begin{figure}
    \centering
    \input{figures/func_HDlf}
    \caption{Flowchart of the function HDlf}
    \label{func_HDlf}
\end{figure}
\subsection{Function curvature} \label{func_Curvature_sec}
The calculation of the curvature in the pit of the dale is taken from the annex B of ISO21910-2 \cite{21920-2} with the estimation of derivatives by a polynomial of sixth degree, which is used in the majority of commercially available software packages. This requires integer indices. If the index is interpolated due to a plateau, the curvature is calculated for the two surrounding indices and averaged. The algorithm was also delegated to a function here. However, we have omitted a flowchart, as the code is lengthy and primarily corresponds to the calculation rule in ISO 21920-2 \cite{21920-2}. The Matlab code can be found in the \ref{feature_attribute_appendix}.
\subsection{Function feature\_characterization} \label{func_FC_sec}
The function \textbf{feature\_characterization}, shown in figure \ref{func_FC}, primarily serves as a parser that decomposes inputs and, if necessary, converts them into suitable formats for the two subfunctions \textbf{watershed\_segmentation} and \textbf{feature\_parameter}. To maintain the translatability of the shown algorithm, the partially implemented parser functions were
deliberately omitted. In the case of feature type and feature attribute, no decomposition is required. These can be passed directly to the functions as $FT$ and $AT$. Other entries may consist of multiple components, which should always be separated by a space. To this end, functions or methods that are implemented in most programming languages are utilized. These include replacing strings (Matlab: strrep), splitting strings by defined delimiters (Matlab: split), and converting strings to float (Matlab: str2double). 
In the context of pruning threshold $TH$ and significant features, the nesting index $NI_{\mathrm{sig}}$, can be specified as a percentage. To ensure the presence of a space between the value and the percentage sign, “\%” is replaced with “ \%”. If “\%” is not present, the string remains unchanged. If the percentage sign initially contained a space, this results in the presence of double spaces within the string. To decompose the entries, the split function is utilized, specifying spaces as the delimiter. Subsequently, the number of substrings, denoted as $N$, is determined. In instances where a double space precedes “\%”, $N$ equals 4 instead of 3.
In the case of the $pruning$ entry, the first word corresponds to the pruning type $PT$. The second part is merely a (real) number and is thus converted from a string to a number and adopted as the threshold $TH$. If $pruning = "\mathrm{None}"$, there is no second part of the string. Then, str2double is applied to the first entry, achieved by $\mathrm{min}(2,\,N)$ as index. The result of str2double on $"\mathrm{None}"$ leads to $TH=\mathrm{NaN}$ (Not a Number), which, however, poses no problem for the further process, since no $TH$ is used for $PT = "\mathrm{None}"$. Furthermore, there is the possibility to determine the optimal threshold. This is the case when $N=2$ and the second entry is $"\mathrm{opt}"$.
If $N \geq 3$, it means that TH was specified in percentage. If this is the case, the already determined value of $TH$ is divided by 100 and multiplied by $Rz$ if $PT="\mathrm{Wolfprune}"$ or multiplied by the evaluation length $l_\mathrm{e}$ if $PT="\mathrm{Width}"$, as defined in ISO 21920-2 \cite{21920-2}. The other pruning types do not provide such a percentage value.
\begin{figure}
    \centering
    \input{figures/func_FC}
    \caption{Flowchart of the function feature\_characterization}
    \label{func_FC}
\end{figure}
Following the same scheme, the $significant$ entry is divided. Here, the first entry corresponds to $F_{\mathrm{sig}}$. If a second entry is present, it corresponds to $NI_{\mathrm{sig}}$. Otherwise, in the case of $F_{\mathrm{sig}}="\mathrm{All}"$, $NI_{\mathrm{sig}}=\mathrm{NaN}$. If $NI_{\mathrm{sig}}$ was specified in percentage, the value for the inverse material ratio $Rcm(p)$ is used. The result of $Rcm(p)$ is the height measured from the highest point of the profile. For the function \textbf{feature\_parameter}, the value must be translated into an absolute height by adding it to $\mathrm{max}(\bi{z})$.
For dividing stats, a percentage input does not need to be considered. Here, the first entry corresponds to $A_\mathrm{stats}$, and $v_\mathrm{stats}$ corresponds to the second entry, if present.
With these defined variables, the functions \textbf{watershed\_segmentation} and \textbf{feature\_parameter} can now be executed, thus determining the corresponding feature parameter $x_\mathrm{FC}$. Finally, the parameters for feature characterization, as well as partial results thereof ($\bi{attr}$), are stored in a structured array named ${META}$. This serves both for traceability and partly for visualization needs ($F_{\mathrm{sig}}$, $NI_{\mathrm{sig}}$).

\subsection{Function optimal\_periodicity} \label{func_optimal_TH_sec}

Adapting the threshold value to the specific needs and circumstances can be a challenging process that requires a certain degree of familiarity and experience. A priori knowledge can be helpful here. Typically, in a turning process, for example, the feed rate can be a guide value, which should correspond to the dale local width. However, the threshold value should not be selected too close to the feed rate. Otherwise, two features will be incorrectly combined due to deviations. On the other hand, if the threshold is too far away from the feed, oversegmentation may occur. In addition, other variables can also be used as reference points that are derived directly from the production parameters or by simulating the production process. An alternative approach is implemented using the \textbf{optimal\_periodicity} function (see figure \ref{func_optimal_TH}). The aim of this function is to determine the threshold value at which the motifs are as equal as possible, i.e., the periodicity is optimized. Potential applications include deterministic surfaces such as turned or micro-structured surfaces. It is always recommended that pruning be based on local volume or local developed length, as these incorporate both width and height. It is possible for features to be very similar in height but differ significantly in width, or vice versa.

The function is based on the approach of Blateyron \textit{optimal limit A}, which he applied to the motif method \cite{Blateyron_2021}. The variable $Q$ is used as a measure of periodicity, which is equal to the mean value of the attribute values of the motifs in relation to their standard deviation. First, a minimum value for the periodicity $Q_{\mathrm{min}}$ is defined. Depending on the attribute type, the default value is set as the limit value, which is then overwritten if $Q_{\mathrm{min}}$ exceeded. Now the minimum motif is pruned until there are only three motifs left. At each iteration, Q is calculated and checked to see if it is greater than $Q_{\mathrm{min}}$. If so, the threshold is overwritten with the minimum attribute value and $Q_{\mathrm{min}}$ updated to the currently calculated $Q$.

\begin{figure}
    \centering
    \input{figures/func_optimal_periodicity_isolated}
    \caption{Flowchart of the function optimal\_periodicity}
    \label{func_optimal_TH}
\end{figure}
\onecolumn
\section{Matlab implementation} \label{Matlab}
\subsection{Watershed segmentation} \label{watershed_segmentation}
\matlabscript{Appendix/watershed_segmentation}{}
\clearpage
\subsection{feature parameter} \label{feature_parameter_appendix}
\matlabscript{Appendix/feature_parameter}{}
\subsection{feature attribute} \label{feature_attribute_appendix}
\matlabscript{Appendix/feature_attribute}{}
\subsection{feature characterization} \label{feature_characterization_appendix}
\matlabscript{Appendix/feature_characterization}{}

\end{document}

%% file: style/flowchart_settings.tex
\usepackage{tikz}
\usetikzlibrary{shapes.geometric, arrows, positioning}
\tikzset{
  startstop/.style={rectangle, rounded corners, minimum height=0.6cm, text centered, draw=black, align=center},
  io/.style={trapezium, trapezium left angle=70, trapezium right angle=110, text centered, draw=black, align=center},
  process/.style={rectangle, minimum width=3cm, minimum height=0.6cm, text centered, draw=black, align=center},
  decision/.style={diamond, minimum width=3cm, minimum height=0.6cm, text centered, draw=black,aspect=2.5,align=center},
  arrow/.style={thick,->,>=stealth},
  }

\newcommand{\V}{1.5pt}      
\newcommand{\Sn}{0.4}      
\newcommand{\Sns}{0.3}      

%% file: style/matlab_code_settings.tex
\usepackage{listings}

\definecolor{MyDarkGreen}{rgb}{0.0,0.4,0.0}
\definecolor{Blue}{rgb}{0.0,0,1.0}
\definecolor{Purple}{rgb}{0.5,0.0,0.5}

\newcommand{\matlabscript}[2]
  {\begin{itemize}\item[]\lstinputlisting[caption=#2,label=#1]{#1.m}\end{itemize}}

%% file: figures/step1.tex
\begin{tikzpicture}[node distance=2cm]

\node (start) [startstop] {
start\\step 1
};

\node (checkfeature) [decision, below=\Sns of start,align=center] {
 $FT \in \{"\mathrm{H}", "\mathrm{P}"\}$
};

\node (invertz) [process, below=\Sns of checkfeature] {
  $\bi{z} := -\bi{z}$
};

\node (findineq) [process, below=\Sns of invertz] {
  $\bi{i_{\mathbf{Neq}}} := [1, \mathrm{find}([{z}_{i} \neq {z}_{i+1}]\,\mathrm{with}\, i=1,2,...,n-1)]$\\[\V]
  $n_{Neq}:=\mathrm{length}(\bi{i_{\mathbf{Neq}}})$\\[\V]
  $\bi{dz} :=z_{i_{\mathrm{Neq}_i}+1} - z_{i_{\mathrm{Neq}_i}}\,\mathrm{with}\, i=1,2,...,n_{\mathrm{Neq}}-1$\\[\V]
  $\bi{s} := \mathrm{sign}(\bi{dz})$\\[\V]
  $\bi{ds} := s_{i+1} - s_{i}\,\mathrm{with}\, i=1,2,...,\mathrm{length}(\bi{s})-1$\\[\V]
  $\bi{i_{\mathbf{p}}} := 1 + \mathrm{find}([\bi{ds} = -2])$\\[\V]
  $\bi{i_{\mathbf{v}}} := 1 + \mathrm{find}([\bi{ds} = 2])$\\[\V]
  $\bi{i_\mathbf{pv}} := [\bi{i_{\mathbf{Neq}_{i_\mathbf{p}}}}, \bi{i_{\mathbf{Neq}_{i_\mathbf{v}}}}]$
};

\node (plateau) [process, below=\Sns of findineq, minimum width=7.5cm] {
  $ \bi{k} := \mathrm{find}([\bi{z_{\bi{i}_{\mathrm{pv}}}} = \bi{z_{\bi{i}_{\mathrm{pv}}+1}}])$\\[\V]
  $\mathrm{for\, each\,}k$\\
  \\ \\
};

\node (each_plateau) [process, below=-1.25cm of plateau] {
    $n_{\mathrm{plateau}} := \mathrm{find}([\bi{z_{i_{\mathbf{pv}_{k+i}}}} \neq {z}_{{i}_{\mathrm{pv}_k}}],"\mathrm{First}")-1$\\[\V]
    $i_{\mathrm{pv}_k} := i_{\mathrm{pv}_k} + (n_{\mathrm{plateau}} - 1)/2$
};

\node (separate) [process, below=\Sns of plateau, align=center] {
  ${\bi{i_\mathbf{p}}} := i_{\mathrm{pv}_i} \,\mathrm{with}\, i =1,2,..., \mathrm{length}(\bi{i}_{\mathrm{p}})$ \\[\V]
  ${\bi{i_\mathbf{v}}} := i_{\mathrm{pv}_i} \,\mathrm{with}\, i = \mathrm{length}(\bi{i_{\mathbf{p}}})+1,...,\mathrm{length}(\bi{i_{\mathbf{pv}}})$
};

\node (end) [startstop, below=\Sns of separate] {
  end\\ step 1
};

\draw [arrow] (start) -- (checkfeature);
\draw [arrow] (checkfeature) -- node[anchor=west] {yes} (invertz);
\draw [arrow] (invertz) -- (findineq);
\draw [arrow] (checkfeature) -- node[anchor=south,xshift=-0.1cm] {no}  ++(3,0) |- ++(-3,-1.65-\Sns);
\draw [arrow] (findineq) -- (plateau);
\draw [arrow] (plateau) -- (separate);
\draw [arrow] (separate) -- (end);

\end{tikzpicture}

%% file: figures/step2.tex
\begin{tikzpicture}[node distance=2cm]
\node (start) [startstop] {
start\\step 2
};

\node (endpoints) [process, below=\Sns of start] 
{
if $i_{\mathrm{v}_1} < i_{\mathrm{p}_1}$ delete $i_{\mathrm{v}_1}$\\[\V]
if $i_{\mathrm{v}_{\mathrm{length}(\bi{i}_\mathbf{v})}} > i_{\mathrm{p}_{\mathrm{length}(\bi{i}_\mathbf{v})}}$ delete $i_{\mathrm{v}_{\mathrm{length}(\bi{i}_\mathbf{v})}}$\\[\V]
$n_M := \mathrm{length}(\bi{i_\mathbf{v}})$\\[\V]
$k:=1$
};

\node (loop) [decision, below=\Sns of endpoints] {
  $k \leq n_\mathrm{M}$
};
\node (foreach) [process, below=\Sns of loop] {
  $[{i}_{\mathrm{lp}},\, {i}_{\mathrm{hp}}] := \mathrm{get\_ilp\_ihp}(\bi{z}, [i_{\mathrm{p}_k}, i_{\mathrm{p}_{k+1}}])$\\[\V]
  $\bi{i_{\mathbf{hi}}} := \mathrm{height\_intersections}(\bi{z}, {i}_{\mathrm{lp}}, {i}_{\mathrm{hp}})$\\[\V]
  ${M}_k := \{i_{\mathrm{v}_k},i_{\mathrm{lp}},i_{\mathrm{hp}},\bi{i_{\mathbf{hi}}},1\}$\\[\V]
  $k:=k+1$
};

\node (end) [startstop, below=\Sns of foreach] {
  end\\step 2
};

\draw [arrow] (start) -- (endpoints);
\draw [arrow] (endpoints) -- (loop);
\draw [arrow] (loop) -- node[anchor=west] {yes} (foreach);
\draw [arrow] (loop) -- node[anchor=south,xshift=.8cm] {no} ++(-3.7,0) |- (end);
\draw [arrow] (foreach) -- ++(3.7,0) |- (loop);

\end{tikzpicture}

%% file: figures/step3.tex
  
\begin{tikzpicture}[node distance=2cm]
\newcommand{\B}{4}

\node (start) [startstop] {
start\\step 3
};

\node (PTnone) [decision, below=\Sns of start] {
$PT \neq "\mathrm{None}"$
};

\node (attr) [process, below=\Sns of PTnone] {
$\bi{attr} := \mathrm{feature\_attribute} (\bi{z}, dx, {M}, PT)$
};
\node (while) [decision,below=\Sns of attr] {
$\mathrm{min}(\bi{attr})<TH$
};
\node (delete) [process, below=\Sns of while] {
$r_\mathrm{min} := \mathrm{min}\{i \,|\, \bi{attr} = \mathrm{min}(\bi{attr})\}$\\[\V]
${M}_\mathrm{min} := {M}_{r_\mathrm{min}}$\\[\V]
delete ${M}_{r_\mathrm{min}}$, delete $\bi{attr}_{r_\mathrm{min}}$\\[\V]
$n_\mathrm{M}:=n_\mathrm{M}-1$\\[\V]
$dir=\mathrm{sign}({M}_\mathrm{min}.i_{\mathrm{lp}}-{M}_\mathrm{min}.i_v)$\\[\V]
$r_\mathrm{U} := \cases{r_\mathrm{min}-1, & $ dir=-1$\\
r_\mathrm{min}, & $\mathrm{otherwise}$}$
};

\node (case1) [decision, aspect=3, below=\Sns of delete] {
 $r_\mathrm{U} = 0 \lor r_\mathrm{U} > n_\mathrm{M}$
};

\node (case2) [decision, aspect=3, below=\Sns of case1] {
${M}_{r_\mathrm{U}}.i_{\mathrm{lp}}={M}_\mathrm{min}.i_{\mathrm{lp}}$
};

\node (case2yes) [process, below = \Sns of case2] {
$\bi{i_\mathbf{p_{\mathbf{surr}}}} :=[{M}_{r_\mathrm{U}}.i_{\mathrm{hp}}, {M}_\mathrm{min}.i_{\mathrm{hp}}]$\\[\V]
$[i_{\mathrm{lp}},\, i_{\mathrm{hp}}] :=\mathrm{get\_ilp\_ihp}(\bi{z}, \bi{i_\mathbf{p_{\mathbf{surr}}}})$\\[\V]
${M}_{r_\mathrm{U}}.i_{\mathrm{lp}} := i_{\mathrm{lp}}$\\[\V]
${M}_{r_\mathrm{U}}.i_{\mathrm{hp}} := i_{\mathrm{hp}}$
};

\node (case3yes) [process, below= \Sns of case2yes] {
${M}_{r_\mathrm{U}}.i_{\mathrm{hp}} := {M}_\mathrm{min}.i_{\mathrm{hp}}$
};

\node (case31) [decision, aspect=3,below=\Sns of case3yes] {
$z_{{M}_{r_\mathrm{U}}.i_{\mathrm{lp}}} \leq z_{{M}_\mathrm{min}.i_\mathrm{v}}$
};
\node (update) [process, below=\Sns of case31] {
${M}_{r_\mathrm{U}}.\bi{i_{\mathbf{hi}}} := \mathrm{height\_intersections}(\bi{z}, i_{\mathrm{lp}},i_{\mathrm{hp}})$\\
$\bi{attr}_{r_\mathrm{U}} := \mathrm{feature\_attribute}(\bi{z}, dx, {M}_{r_\mathrm{U}}, PT)$
};

\node (end) [startstop, left = 1.2cm of start] {
end\\step 3
};

\draw [arrow] (start) -- (PTnone);
\draw [arrow] (PTnone) -- node[anchor=west] {yes} (attr);
\draw [arrow] (attr) -- (while);
\draw [arrow] (while) -- node[anchor=west] {yes} (delete);
\draw [arrow] (delete) -- (case1);

\draw [arrow] (while) -- node[anchor=south, align=center,xshift=0.5cm] {no} ++(-\B,0) |- (end);
\draw [arrow] (PTnone) -- ++(-\B,0);

\draw [arrow] (case1) -- node[anchor=west] {no} (case2);
\draw [arrow] (case1) -- node[anchor=south, align=center,xshift=-0.3cm] {yes \\(case 1)} ++(\B,0);

\draw [arrow] (case2) -- node[anchor=west, align=center,] {yes (case 2)} (case2yes);
\draw [arrow] (case2) -- ++(3,0) node[anchor=south, align=center] {no \\(case 3)} |- (case3yes);
\draw [arrow] (case2yes) -- ++(-3,0) |- ++(3,-4.03);

\draw [arrow] (case3yes) -- (case31);
\draw [arrow] (case31) -- node[anchor=west, align=center,] {no (case 3.2)} (update);  
\draw [arrow] (case31) -- node[anchor=south, align=center,xshift=-0.1cm,yshift=0.0cm] {yes \\(case 3.1)} ++(\B,0);

\draw [arrow] (update) |- ++(\B,-.7) |- (while);

\end{tikzpicture}

%% file: figures/step4.tex
\begin{tikzpicture}[node distance=2cm]
\newcommand{\B}{3.95}
\node (start) [startstop] {
  start\\step 4
};

\node (INsig) [process, below = \Sns of start, minimum width=0cm] {
    $\bi{I_\mathbf{Nsig}}:=[\,]$\\
    $n_\mathrm{M}=\mathrm{length(M)}$
};

\node (Fsig) [decision, below = \Sns of INsig, aspect = 3] {
$\mathrm{switch}\,F_{\mathrm{sig}}$
};

\node (TopBot) [process, below = \Sn of Fsig] {
    $NI_{\mathrm{sig}}:=\mathrm{min}(NI_{\mathrm{sig}},n_\mathrm{M})$\\[\V]
    $\bi{h} := \mathrm{feature\_attribute} (\bi{z}, dx, {M}, n_\mathrm{M}, "\mathrm{PVh}")$\\[\V]
    $\mathrm{sort\,} \bi{h} \mathrm{\,such\, that\,} h_{i_1} \geq h_{i_2} \geq.... \geq h_{i_{n_\mathrm{M}}}$\\[\V]
    $\bi{I_\mathbf{Nsig}}:=[i_{NI_{\mathrm{sig}}+1}, i_{NI_{\mathrm{sig}}+2}, ..., i_{n_\mathrm{M}}]$
};

\node (zlp) [process, below = \Sns of TopBot] {
  $\bi{z_\mathbf{lp}} := {z}_{\mathrm{floor}({M}_i.i_{\mathrm{lp}})} \, \mathrm{with}\, i = 1,2,...,n_\mathrm{M}$\\[\V]
  $FTI:=\mathrm{sign}(z_{\mathrm{floor}({M}_1.i_{\mathrm{lp}})} - z_{\mathrm{floor}({M}_1.i_\mathrm{v})})$
};

\node (Open) [decision, below = \Sns of zlp, aspect = 3] {
  $F_{\mathrm{sig}}="\mathrm{Open}"$
};

\node (Openyes) [process, below = \Sn of Open] {
$\bi{I_\mathbf{Nsig}} := \mathrm{find}([FTI\cdot \bi{z_{\mathbf{lp}}} > FTI\cdot NI_{\mathrm{sig}}])$
};

\node (Openno) [process, below = \Sns of Openyes] {
$\bi{z}_{\mathbf{v}}:= {z}_{\mathrm{floor}({M}_i.i_\mathrm{v})} \, \mathrm{with}\, i = 1,2,...,n_\mathrm{M}$\\[\V]
$\bi{I_\mathbf{Nsig}} := \mathrm{find}([FTI\cdot \bi{z_{\mathbf{lp}}} < FTI\cdot NI_{\mathrm{sig}} \lor ...$\\$FTI\cdot \bi{z_\mathbf{v}} > FTI\cdot NI_{\mathrm{sig}}])$
};

\node (setsigzero) [process, below = \Sns of Openno] {
${M}_i.sig:=0$ for all $i \in \bi{I_\mathbf{Nsig}}$
};

\node (end) [startstop, below = \Sns of setsigzero] {
  end\\ step 4
};

\draw [arrow] (start) -- (INsig);
\draw [arrow] (INsig) -- (Fsig);
\draw [arrow] (Fsig) node[anchor=south,yshift=0.2cm,xshift=1.8cm] {$\mathrm{"All"}$} -- ++(\B,0) |- (setsigzero);
\draw [arrow] (Fsig) node[anchor=south,yshift=0.2cm,xshift=-2.25cm] {$\mathrm{"Open"} \lor \mathrm{"Closed"}$} -- ++(-\B,0) |- (zlp);
\draw [arrow] (zlp) -- (Open);
\draw [arrow] (Open) -- node[anchor=east] {yes}(Openyes);
\draw [arrow] (Open) node[anchor=south,xshift=-2.2cm] {no} -- ++(-\B,0) |- (Openno);
\draw [arrow] (Openno) -- ++(\B,0);
\draw [arrow] (Openyes) -- ++(\B,0);

\draw [arrow] (Fsig) node[anchor=west,yshift=-0.8cm] {$\mathrm{"Top"} \lor \mathrm{"Bot"}$} --  (TopBot);

\draw [arrow] (TopBot) -- ++(\B,0);
\draw [arrow] (setsigzero) -- (end);

\end{tikzpicture}

%% file: figures/func_feature_attribute.tex
\begin{tikzpicture}[node distance=2cm]
\newcommand{\B}{2cm}
\newcommand{\Bmid}{2.3cm}
\newcommand{\Bend}{4cm}
\newcommand{\Snss}{0.25cm}
\node (start) [startstop] {
  start\\feature\_attribute
};

\node (input) [io, below=\Sn of start] {
    $\mathrm{input}$ $\bi{z}$, $dx$, ${M}$, $AT$
  };

\node (Isig) [process, below = \Sn of input] {
    $\bi{I_{\mathbf{sig}}} := \mathrm{find}([{M}.sig=1])$
};

\node (Attributetype) [decision, below=\Sn of Isig] 
{
switch\\ AT
};

\node (HDh) [process, below right = -4.5cm and 5.5cm of Attributetype] {
    $\bi{attr}:=|z_{\mathrm{floor}({M}_{\bi{I_{\mathbf{sig}}}}.i_{\mathrm{lp}})}-z_{\mathrm{floor}({M}_{\bi{I_{\mathbf{sig}}}}.i_\mathrm{v})}|$
};

\node (HDw) [process, below=\Snss of HDh] {
    $\bi{attr}:=\mathrm{max}(|dx\cdot({M}_{\bi{I_{\mathbf{sig}}}}.i_{\mathrm{hi}} - {M}_{\bi{I_{\mathbf{sig}}}}.i_{\mathrm{lp}})|)$
};

\node (HDv) [process, below=\Snss of HDw] {
    $\bi{attr}:=\mathrm{HDvf}(\bi{z},dx,{M}_{\bi{I_{\mathbf{sig}}}})$
};

\node (HDl) [process, below=\Snss of HDv] {
    $\bi{attr}:=\mathrm{HDlf}(\bi{z},dx,{M}_{\bi{I_{\mathbf{sig}}}})$
};

\node (PVh) [process, below=\Snss of HDl] {
    $FTI:=\mathrm{sign}({z}_{\mathrm{floor}({M}_1.i_{\mathrm{lp}})} - z_{\mathrm{floor}({M}_1.i_\mathrm{v})})$\\[\V]
    $\bi{attr}:=-FTI\cdot {z}_{\mathrm{floor}({M}_{\bi{I_{\mathbf{sig}}}}.i_\mathrm{v})}$
};

\node (Curvature) [process, below=\Snss of PVh] {
    $\bi{attr} := \mathrm{curvature}(\bi{z}, dx, {M}_{\bi{I_{\mathbf{sig}}}}.i_\mathrm{v})$
};

\node (Count) [process, below=\Snss of Curvature] {
    $\bi{attr}:=1 \mathrm{,\,where}\, {M}_{\bi{I_{\mathbf{sig}}}}.sig = 1$
};

\node (output) [io, below = \Snss of Count] {
  return $\bi{attr}$
};

\node (end) [startstop, below = \Sns of output] {
  end\\feature\_attribute
};

\draw [arrow] (start) -- (input);
\draw [arrow] (input) -- (Isig);
\draw [arrow] (Isig) -- (Attributetype);
\draw [arrow] (Attributetype) -- ++(\Bmid,0) |- node[anchor=south,xshift=\B] {$"\mathrm{Wolfprune}" \lor "\mathrm{HDh}"$} (HDh);
\draw [arrow] (Attributetype) -- ++(\Bmid,0) |- node[anchor=south,xshift=\B] {$"\mathrm{Width}" \lor "\mathrm{HDw}"$} (HDw);
\draw [arrow] (Attributetype) -- ++(\Bmid,0) |- node[anchor=south,xshift=\B] {$"\mathrm{DevLength}" \lor "\mathrm{HDl}"$} (HDl);
\draw [arrow] (Attributetype) -- ++(\Bmid,0) |- node[anchor=south,xshift=\B] {$"\mathrm{VolS}" \lor "\mathrm{HDv}"$} (HDv);
\draw [arrow] (Attributetype) -- ++(\Bmid,0) |- node[anchor=south,xshift=\B] {$"\mathrm{PVh}"$}(PVh);
\draw [arrow] (Attributetype) -- ++(\Bmid,0) |- node[anchor=south,xshift=\B] {$"\mathrm{Curvature}"$}(Curvature);
\draw [arrow] (Attributetype) -- ++(\Bmid,0) |- node[anchor=south,xshift=\B] {$"\mathrm{Count}"$} (Count);

\draw [arrow] (HDh) -- ++(\Bend,0) |- (output);
\draw [arrow] (output) -- (end);
\draw [arrow] (HDw) -- ++(\Bend,0);
\draw [arrow] (HDl) -- ++(\Bend,0);
\draw [arrow] (HDv) -- ++(\Bend,0);
\draw [arrow] (PVh) -- ++(\Bend,0);
\draw [arrow] (Curvature) -- ++(\Bend,0);
\draw [arrow] (Count) -- ++(\Bend,0);

\end{tikzpicture}

%% file: figures/step6.tex
\begin{tikzpicture}[node distance=2cm]
\newcommand{\B}{-0.1cm}
\newcommand{\Y}{0.25cm}
\newcommand{\Bleft}{-4.2cm}
\newcommand{\Bright}{3.5cm}
\newcommand{\Snss}{0.25cm}

\node (start) [startstop] {
  start\\step 6
};

\node (Statstype) [decision, below=\Sns of start] 
{
switch\\ $stats$
};

\node (Mean) [process, below=\Snss of Statstype] {
    $\bi{x_{\mathbf{FC}}} := \frac{1}{n_\mathrm{M}} \sum\limits_{i=1}^{n_\mathrm{M}}{attr}_i$
};

\node (Max) [process, below=\Snss of Mean] {
    $\bi{x_{\mathbf{FC}}} := \mathrm{max}(\bi{attr})$
};

\node (Min) [process, below=\Snss of Max] {
    $\bi{x_{\mathbf{FC}}} := \mathrm{min}(\bi{attr})$
};

\node (Std) [process, below=\Snss of Min] {
    $\bi{x_{\mathbf{FC}}} := \mathrm{std}(\bi{attr})$
};

\node (Perc) [process, below=\Snss of Std] {
    $\bi{x_{\mathbf{FC}}} := \frac{1}{length(attr)} \sum [\bi{attr} > v_{stats}]$
};

\node (Hist) [process, below=\Sns of Perc] {
    $\bi{x_{\mathbf{FC}}} := \mathrm{histogram}(\bi{attr})$
};

\node (Sum) [process, below=\Snss of Hist] {
    $\bi{x_{\mathbf{FC}}} := \sum\limits_{i=1}^{n_\mathrm{M}} {attr}_i$
};

\node (Density) [process, below=\Snss of Sum] {
    $\bi{x_{\mathbf{FC}}} := \frac{1}{dx\cdot \mathrm{length}(\bi{z})}\sum\limits_{i=1}^{n_\mathrm{M}} {attr}_i$
};

\node (end) [startstop, below=\Snss of Density] {
  end\\ step 6
};

\draw [arrow] (start) -- (Statstype);
\draw [arrow] (Statstype) -- ++(\Bleft,0) |- node[anchor=west,xshift=\B,yshift=\Y] {"$\mathrm{Mean}$"} (Mean);
\draw [arrow] (Statstype) -- ++(\Bleft,0) |- node[anchor=west,xshift=\B,yshift=\Y] {"$\mathrm{Max}$"} (Max);
\draw [arrow] (Statstype) -- ++(\Bleft,0) |- node[anchor=west,xshift=\B,yshift=\Y] {"$\mathrm{Min}$"} (Min);
\draw [arrow] (Statstype) -- ++(\Bleft,0) |- node[anchor=west,xshift=\B,yshift=\Y] {"$\mathrm{StdDev}$"} (Std);
\draw [arrow] (Statstype) -- ++(\Bleft,0) |- node[anchor=west,xshift=\B,yshift=\Y] {"$\mathrm{Perc}$"}(Perc);
\draw [arrow] (Statstype) -- ++(\Bleft,0) |- node[anchor=west,xshift=\B,yshift=\Y] {"$\mathrm{Hist}$"}(Hist);
\draw [arrow] (Statstype) -- ++(\Bleft,0) |- node[anchor=west,xshift=\B,yshift=\Y] {"$\mathrm{Sum}$"} (Sum);
\draw [arrow] (Statstype) -- ++(\Bleft,0) |- node[anchor=west,xshift=\B,yshift=\Y] {"$\mathrm{Density}$"} (Density);

\draw [arrow] (Mean) -- ++(\Bright,0) |- (end);
\draw [arrow] (Max) -- ++(\Bright,0);
\draw [arrow] (Std) -- ++(\Bright,0);
\draw [arrow] (Min) -- ++(\Bright,0);
\draw [arrow] (Perc) -- ++(\Bright,0);
\draw [arrow] (Hist) -- ++(\Bright,0);
\draw [arrow] (Sum) -- ++(\Bright,0);
\draw [arrow] (Density) -- ++(\Bright,0);

\end{tikzpicture}

%% file: figures/func_find.tex
\begin{tikzpicture}[node distance=2cm]

\node (start) [startstop] {
  start \\ find
};

\node (input) [io, below=\Sns of start] {
input $\bi{A}$, $FirstOnly$
};

\node (exist) [decision, below=\Sns of input] 
{
$FirstOnly$ exist?
};

\node (existno) [process, below=\Sns of exist] {
    $FirstOnly:=0$
};

\node (initial) [process, below=\Sns of existno] {
    $i:=1$; 
    $\bi{i_{\mathbf{N0}}}=[\,]$
};

\node (while) [decision, below=\Sns of initial] 
{
    $i\leq \mathrm{length}(\bi{A})$
};

\node (N0) [decision, below = \Sns of while] 
{
$A_i\neq 0$
};

\node (iterate) [process, right = \Sns of while,minimum width=1cm] 
{
$i:=i+1$
};

\node (N0yes) [process, below = \Sns of N0] {
$\bi{i_{\mathbf{N0}}} := [\bi{i_{\mathbf{N0}}}, i]$
};

\node (first) [decision, below = \Sns of N0yes] 
{
$FirstOnly=0$
};

\node (return) [io, below = \Sns of first] 
{
return $\bi{i_{\mathbf{N0}}}$
};

\node (end) [startstop, below=\Sns of return] 
{end\\find
};

\draw [arrow] (start) -- (input);
\draw [arrow] (input) -- (exist);
\draw [arrow] (exist) -- node[anchor=west] {no}(existno);
\draw [arrow] (existno) -- (initial);
\draw [arrow] (exist) node[anchor=south, xshift=-2.5cm] {yes} -- ++(-2.8,0) |- (initial);
\draw [arrow] (initial) -- (while);
\draw [arrow] (while) -- node[anchor=west] {yes} (N0);
\draw [arrow] (N0) -- node[anchor=west] {yes} (N0yes);
\draw [arrow] (N0yes) -- (first);
\draw [arrow] (first) -- node[anchor=west] {no}(return);
\draw [arrow] (return) -- (end);
\draw [arrow] (first) node[anchor=south, xshift=2.3cm] {yes} -| (iterate);
\draw [arrow] (iterate) -- (while);
\draw [arrow] (N0) node[anchor=south, xshift=1.7cm] {no} -- ++(3.2,0);

\end{tikzpicture}

%% file: figures/func_get_ilp_ihp.tex
\begin{tikzpicture}[node distance=2cm]

\node (start) [startstop] {
  start\\ get\_ilp\_ihp
};

\node (input) [io, below=\Sns of start] {
  $\mathrm{input}$ $\bi{z}$,
  $\bi{i_\mathbf{p_{\mathbf{surr}}}}$
};

\node (floor) [process, below=\Sns of input] {
  $\bi{z_{\mathbf{surr}}}=[{z}_{\mathrm{floor}({i}_{\mathrm{p}_{\mathrm{surr}_1}})}, \, {z}_{\mathrm{floor}({i}_{\mathrm{p}_{\mathrm{surr}_2}})}]$\\[\V]
  $I := \mathrm{min}\{i \,|\, \bi{z_{\mathbf{surr}}} = \mathrm{min}(\bi{z_{\mathbf{surr}}})\}$\\[\V]
  ${i}_{\mathrm{lp}}={i}_{\mathrm{p}_{\mathrm{surr}_I}}$\\[\V] 
  ${i}_{\mathrm{hp}}={i}_{\mathrm{p}_{\mathrm{surr}_{3-I}}}$
};

\node (output) [io, below=\Sns of floor] {
 $\mathrm{return}$ ${i}_{\mathrm{lp}}$, ${i}_{\mathrm{hp}}$
};

\node (end) [startstop, below=\Sns of output] 
{end\\ get\_ilp\_ihp
};

\draw [arrow] (start) -- (input);
\draw [arrow] (input) -- (floor);
\draw [arrow] (floor) -- (output);
\draw [arrow] (output) -- (end);

\end{tikzpicture}

%% file: figures/func_height_intersections.tex
\begin{tikzpicture}[node distance=2cm]
\newcommand{\B}{3.9}
\node (start) [startstop] {
  start\\height\_intersections
};

\node (input) [io, below=\Sns of start] {
  $\mathrm{input}$ $\bi{z}$,
  $i_{\mathrm{lp}}$,
  $i_{\mathrm{hp}}$
};

\node (init) [process, below=\Sns of input] 
{
  $\bi{i_{\mathbf{hi}}}:=[\,]$; $dir:=\mathrm{sign}(i_{\mathrm{hp}} - i_{\mathrm{lp}})$\\[\V]
  $i_{\mathrm{lp}}:=\mathrm{round}(i_{\mathrm{lp}})$; $i_{\mathrm{hp}}:=\mathrm{round}(i_{\mathrm{hp}})$; $z_{\mathrm{lp}} := z_{i_{\mathrm{lp}}}$\\[\V]
  $j := i_{\mathrm{lp}}+dir\cdot(\mathrm{find}(\bi{z}(i_{\mathrm{lp}}:dir:i_{\mathrm{hp}}) \neq z_{\mathrm{lp}},1)-1)$
};

\node (while) [decision, below=\Sns of init, aspect=2.6]{
$j \neq i_{\mathrm{hp}}$
};

\node (if) [decision, below=\Sns of while, aspect=2.3]{
${z}_j < z_{\mathrm{lp}} \land {z}_{j+dir} \geq z_{\mathrm{lp}}$ \\ 
$\lor$ \\
${z}_j \geq z_{\mathrm{lp}} \land {z}_{j+dir} < z_{\mathrm{lp}}$
};

\node (ihi) [process, below=\Sns of if] 
{
    $\bi{i_{\mathbf{hi}}}:=[\bi{i_{\mathbf{hi}}}; j + dir \cdot(z_{\mathrm{lp}} - {z}_j)/({z}_{j + dir} - {z}_j]$
};

\node (iterate) [process, below=\Sns of ihi] 
{
    $j:=j+dir$
};

\node (output) [io, below=\Sns of iterate] {
    $\mathrm{return}$ $\bi{i_{\mathbf{hi}}}$
};

\node (end) [startstop, below=\Sns of output] 
{end\\height\_intersections
};

\draw [arrow] (start) -- (input);
\draw [arrow] (input) -- (init);
\draw [arrow] (init) -- (while);
\draw [arrow] (while) -- node[anchor=west, yshift=-0cm] {yes} (if);
\draw [arrow] (if) -- node[anchor=west, yshift=-0cm] {yes} (ihi);
\draw [arrow] (ihi) -- (iterate);
\draw [arrow] (output) -- (end);

\draw [arrow] (iterate) -- ++(-\B,0) |- (while);
\draw [arrow] (while) -- node[anchor=south, xshift=-.8cm] {no} ++(\B,0) |- (output);
\draw [arrow] (if) -- node[anchor=south, xshift=-0cm] {no} ++(3.65,0) |- (iterate);

\end{tikzpicture}

%% file: figures/func_HDvf.tex
\begin{tikzpicture}[node distance=2cm]
\newcommand{\R}{4cm}
\newcommand{\B}{3.95}

\node (start) [startstop] {
  start\\HDvf
};

\node (input) [io, below=\Sns of start] {
    $\mathrm{input}$ $\bi{z}$, $dx$, ${M}_\mathrm{r}$
  };
  
\node (init) [process, below=\Sns of input] {
    $z_{\mathrm{lp}} := {z}_{\mathrm{floor}({M}_\mathrm{r}.i_{\mathrm{lp}})}$\\[\V]
    $dir:=\mathrm{sign}({M}_\mathrm{r}.i_{\mathrm{hp}} - {M}_\mathrm{r}.i_{\mathrm{lp}});$\\[\V]
    $A := 0$; $i := 1$\\[\V]
    $\bi{i_{\mathbf{hi}}} := [{M}_\mathrm{r}.i_{\mathrm{lp}}; {M}_\mathrm{r}.\bi{i_{\mathbf{hi}}}]$
};

\node (while) [decision, below=\Sns of init] {
    $i \leq \mathrm{length}(\bi{i_{\mathbf{hi}}})$
};

\node (loop) [process, below=\Sn of while] {
    $i_1 := |\mathrm{ceil}(dir\cdot i_{{{\mathrm{hi}}}_i})|$\\[\V]
    $i_2 := |\mathrm{floor}(dir\cdot i_{{{\mathrm{hi}}}_{i+1}})|$\\[\V]
    $\bi{x_\mathrm{f}} := dx\cdot [i_{{\mathrm{hi}}_i}; (i_1:dir:i_2); i_{{\mathrm{hi}}_{i+1}}]$\\[\V]
    $\bi{z_\mathrm{f}} := [z_{\mathrm{lp}}; \bi{z}(i_1:dir:i_2); z_{\mathrm{lp}}]$\\[\V]
    $A := A + |\mathrm{trapz}(\bi{x_\mathrm{f}}, \bi{z_\mathrm{f}}-z_{\mathrm{lp}}|$\\[\V]
    $i:=i+2$
};

\node (output) [io, below=\Sns of loop] {
return $HDv := A/(\mathrm{length}(\bi{z})\cdot dx)$
};

\node (end) [startstop, below =\Sns of output] {
  end\\HDvf
};

\draw [arrow] (start) -- (input);
\draw [arrow] (input) -- (init);
\draw [arrow] (init) -- (while);
\draw [arrow] (while) -- node[anchor=west, yshift=-0cm] {yes} (loop);
\draw [arrow] (loop) |- ++(\B,-1.62) |- (while);
\draw [arrow] (while) -- node[anchor=south, xshift=0.7cm] {no} ++(-\B,0) |- (output);
\draw [arrow] (output) -- (end);

\end{tikzpicture}

%% file: figures/func_HDlf.tex
\begin{tikzpicture}[node distance=2cm]
\newcommand{\R}{4cm}
\newcommand{\B}{3.95}

\node (start) [startstop] {
  start\\HDlf
};

\node (input) [io, below=\Sns of start] {
    $\mathrm{input}$ $\bi{z}$, $dx$, ${M}_\mathrm{r}$
  };
  
\node (init) [process, below=\Sns of input] {
    $z_{\mathrm{lp}} := z_{\mathrm{floor}({M}_\mathrm{r}.i_{\mathrm{lp}})}$\\[\V]
    $dir:=\mathrm{sign}({M}_\mathrm{r}.i_{\mathrm{hp}} - {M}_\mathrm{r}.i_{\mathrm{lp}})$\\[\V]
    $i_{{\mathrm{hi}}_{\mathrm{end}}} := {M}_\mathrm{r}.i_{{\mathrm{hi}}_{\mathrm{length}({M}_\mathrm{r}.\bi{i}_{\mathbf{hi}})}}$\\[\V]
    $i_1 := |\mathrm{ceil}(dir\cdot {M}_\mathrm{r}.i_{\mathrm{lp}})|$\\[\V]
    $i_2 := |\mathrm{floor}(dir\cdot i_{{\mathrm{hi}}_{\mathrm{end}}})|$\\[\V]
    $\bi{z_\mathrm{f}} := [\bi{z}(i_1:dir:i_2)]$\\[\V]
    $\bi{dz} := z_{\mathrm{f}_{i+1}} - z_{\mathrm{f}_i}\,\mathrm{with}\, i=1,2,...,\mathrm{length}(\bi{z}_\mathbf{f})-1$\\[\V]
    $ HDl:=\sum_{i=1}^{\mathrm{length}(\bi{dz})}\sqrt{dx^2 + dz_i^2}\,...$\\[\V]
    $... + \mathrm{mod}(i_1, 1)\cdot dx...$\\[\V]$...+\sqrt{((i_{{\mathrm{hi}}_{\mathrm{end}}}-i_2)\cdot dx)^2+(z_\mathrm{lp}-z_{i_2})^2}$
};

\node (output) [io, below=\Sns of init] {
return $HDl$
};

\node (end) [startstop, below =\Sns of output] {
  end\\HDlf
};

\draw [arrow] (start) -- (input);
\draw [arrow] (input) -- (init);
\draw [arrow] (init) -- (output);
\draw [arrow] (output) -- (end);
\end{tikzpicture}

%% file: figures/func_FC.tex
\begin{tikzpicture}[node distance=2cm]
\newcommand{\R}{4cm}
\newcommand{\B}{3.95}
\node (start) [startstop] {
  start feature\_characterization
};

\node (input) [io, below=\Sns of start] {
    $\mathrm{input}$ $\bi{z}$, $dx$, $FT$, $pruning$,\\ $significant$, $AT$, $stats$
  };
\node (split1) [process, below = \Sns of input] {
    $str:=$ replace "\%" with " \%" in $pruning$\\[\V]
    $\bi{str} :=$ split $str$ with delimeter "\," \\[\V]
    $N := \mathrm{length}(\bi{str})$\\[\V]
    $PT := {str}_1$; $TH := \mathrm{str2double}({str}_{min(2,N)})$
};

\node (optcheck) [decision, below=\Sns of split1,aspect=4] {
    $N=2 \land str_2="\mathrm{opt}"$
};
\node (THopt) [process, below = \Sn of optcheck] {
    $TH:="\mathrm{opt}"$
};
\node (perccheck1) [decision, below=\Sn of THopt] {
    $N \geq 3$
};
\node (THperc) [process, below = \Sn of perccheck1] {
    $TH := \cases{0.01\cdot{TH}\cdot Rz,&\hspace{-14pt}$PT="\mathrm{Wolfprune}"$\\
    0.01\cdot{TH}\cdot l_e,&\hspace{-14pt}$PT="\mathrm{Width}"$}$
};
\node (split2) [process, below = \Sn of THperc] {
    $str:=$ replace "\%" with " \%" in $significant$\\[\V]
    $\bi{str} :=$ split $str$ with delimeter "\," \\[\V]
    $N := \mathrm{length}(\bi{str})$\\[\V]
    $F_{sig} := {str}_1$; $NI_{\mathrm{sig}} := \mathrm{str2double}({str}_{min(2,N)}$
};
\node (perccheck2) [decision, below=\Sns of split2] {
    $N\geq3$
};
\node (NIperc) [process, below = \Sn of perccheck2] {
    $TH := \mathrm{max}(\bi{z}) + Rcm(NI_{\mathrm{sig}})$
};
\node (split3) [process, below = \Sn of NIperc] {
    $\bi{str} :=$ split $str$ with delimeter "\," in $stats$\\[\V]
    $N = \mathrm{length}(\bi{str})$\\[\V]
    $A_{stats} := {str}_1$; $v_{\mathrm{stats}} := \mathrm{str2double}({str}_{N})$
};
\node (fp) [process, below = \Sns of split3] {
    ${M}:=\mathrm{watershed\_segmentation}(\bi{z}, dx, FT, PT, TH)$\\[\V]
    $[\bi{x_{\mathbf{FC}}}, {M}, \bi{attr}]:=\mathrm{feature\_parameter}(\bi{z},dx,{M},...$\\$...F_{\mathrm{sig}},NI_{\mathrm{sig}},AT,A_{\mathrm{stats}},v_{\mathrm{stats}})$\\[\V]
    ${META}:=\{\bi{attr},PT,TH,F_{\mathrm{sig}},NI_{\mathrm{sig}},AT,...$\\$...A_{\mathrm{stats}},v_{\mathrm{stats}}\}$
};

\node (output) [io ,below=\Sns of fp] {
$\mathrm{return}$ $\bi{x_{\mathrm{FC}}}$, ${M}$, ${META}$
};

\node (end) [startstop, below =\Sns of output] {
  end feature\_characterization
};

\draw [arrow] (start) -- (input);
\draw [arrow] (input) -- (split1);
\draw [arrow] (split1) -- (optcheck);
\draw [arrow] (optcheck) -- node[anchor=east] {yes} (THopt);
\draw [arrow] (THopt) -- (perccheck1);
\draw [arrow] (perccheck1) -- node[anchor=east] {yes} (THperc);
\draw [arrow] (THperc) -- (split2);
\draw [arrow] (split2) -- (perccheck2);
\draw [arrow] (perccheck2) -- node[anchor=east] {yes} (NIperc);
\draw [arrow] (NIperc) -- (split3);
\draw [arrow] (split3) -- (fp);
\draw [arrow] (fp) -- (output);
\draw [arrow] (output) -- (end);

\draw [arrow] (optcheck) node[anchor=south,xshift=2.95cm] {no} -- ++(\B,0) |- ++(-\B,-1.9) ;
\draw [arrow] (perccheck1) node[anchor=south,xshift=1.5cm] {no} -- ++(\B,0) |- ++(-\B,-2.4);
\draw [arrow] (perccheck2) node[anchor=south,xshift=1.5cm] {no} -- ++(\B,0) |- ++(-\B,-1.7);

\end{tikzpicture}

%% file: figures/func_optimal_periodicity_isolated.tex
  
\begin{tikzpicture}[node distance=2cm]
\newcommand{\B}{3.9}

  \node (start) [startstop] {
    start \\optimal\_periodicity
  };
  
  \node (input) [io, below=\Sns of start] {
    $\mathrm{input}$ $\bi{z}$, $dx$, $FT$, $PT$ 
  };

  \node (step12) [process, below=\Sns of input] {
    determine peaks and pits (see step 1)\\
    determine motifs (see step 2)
  };
  \node (init) [process, below=\Sns of step12] {
    $Q_{\mathrm{min}}=3$\\
    set default Threshold $TH$ depending on $PT$
  };
  
  \node (while) [decision, below=\Sns of init] {
    $n_\mathrm{M} > 3$
  };

  \node (Q) [process, below = \Sns of while] {
   $Q={\mathrm{mean}(\bi{attr})}/{\mathrm{std}(\bi{attr})}$
  };

  \node (ifQ) [decision, below= \Sns of Q] {
    $Q>Q_{\mathrm{min}}$
  };

  \node (Qyes) [process, below=\Sns of ifQ] {
    $TH=\mathrm{min}(\bi{attr})$\\
    $Q_{\mathrm{min}}:=Q$
  };
  \node (prune) [process, below=\Sns of Qyes] {
    pruning minimal motif (see step 3)
  };

  \node (output) [io,below=\Sns of prune] {
    $\mathrm{return}$ $TH$
  };

  \node (end) [startstop, below = \Sns of output] {
    end \\optimal\_periodicity
  };
  
  \draw [arrow] (start) -- (input);
  \draw [arrow] (input) -- (step12);
  \draw [arrow] (step12) -- (init);
  \draw [arrow] (init) -- (while);
  \draw [arrow] (while) -- node[anchor=west] {yes} (Q);
  \draw [arrow] (Q) -- (ifQ);
  \draw [arrow] (ifQ) -- node[anchor=west] {yes} (Qyes);
  \draw [arrow] (Qyes) -- (prune);
  \draw [arrow] (output) -- (end);

  \draw [arrow] (prune) -- ++(-\B,0) |- (while);
  \draw [arrow] (ifQ) -- node[anchor=south, xshift=-0.2cm] {no} ++(2.5,0) |- ++(-2.5,-2.08);
  \draw [arrow] (while) -- node[anchor=south, xshift=-1cm] {no} ++(\B,0) |- (output);

  \end{tikzpicture}